\documentclass[aps,prb,twocolumn,superscriptaddress,bibnotes]{revtex4-2}
\usepackage{graphicx}
\usepackage{color,amssymb,amsmath}
\usepackage[normalem]{ulem}
\usepackage{hyperref}

\newcommand{\Vsd}{V_{\mathrm{SD}}}
\newcommand{\Bperp}{B_{\perp}}
\newcommand{\Vt}{V_{\mathrm{T}}}
\newcommand{\Ea}{E_{\mathrm{A}}}
\newcommand{\Vtg}{V_{\mathrm{TG}}}
\newcommand{\Vprobe}{V_{\mathrm{Probe}}}
\newcommand{\Idc}{I_{\mathrm{DC}}}

\newcommand{\Vac}{V_{\mathrm{AC}}}
\newcommand{\Gbg}{G_{\mathrm{bg}}}

\newcommand{\Iad}{I_{\mathrm{ad.}}}
\newcommand{\In}{I_{n}}
\newcommand{\dBm}{\mathrm{dBm}}
\newcommand{\Jzero}{J_{0}}
\newcommand{\Ione}{I_{1}}
\newcommand{\Vone}{V_{1}}
\newcommand{\Vtwo}{V_{2}}
\newcommand{\Vmw}{V_{\mathrm{MW}}}

\begin{document}
\title{Microwave-induced conductance replicas in hybrid Josephson junctions without Floquet-Andreev states}

\author{D. Z. Haxell}
\affiliation{IBM Research Europe - Zurich, 8803 R\"uschlikon, Switzerland}

\author{M. Coraiola}
\affiliation{IBM Research Europe - Zurich, 8803 R\"uschlikon, Switzerland}

\author{D. Sabonis}
\affiliation{IBM Research Europe - Zurich, 8803 R\"uschlikon, Switzerland}

\author{M. Hinderling}
\affiliation{IBM Research Europe - Zurich, 8803 R\"uschlikon, Switzerland}

\author{S. C. ten Kate}
\affiliation{IBM Research Europe - Zurich, 8803 R\"uschlikon, Switzerland}

\author{E. Cheah}
\affiliation{Solid State Laboratory, ETH Z\"urich, 8093 Z\"urich, Switzerland}

\author{F. Krizek}
\affiliation{IBM Research Europe - Zurich, 8803 R\"uschlikon, Switzerland}
\affiliation{Solid State Laboratory, ETH Z\"urich, 8093 Z\"urich, Switzerland}

\author{R. Schott}
\affiliation{Solid State Laboratory, ETH Z\"urich, 8093 Z\"urich, Switzerland}

\author{W. Wegscheider}
\affiliation{Solid State Laboratory, ETH Z\"urich, 8093 Z\"urich, Switzerland}

\author{W. Belzig}
\affiliation{Fachbereich Physik, Universit\"at Konstanz, D-78457 Konstanz, Germany}

\author{J. C. Cuevas}
\affiliation{Fachbereich Physik, Universit\"at Konstanz, D-78457 Konstanz, Germany}
\affiliation{Departamento de Física Te\'orica de la Materia Condensada and Condensed Matter Physics Center (IFIMAC), Universidad Aut\'onoma de Madrid, E-28049 Madrid, Spain}

\author{F. Nichele}
\email{fni@zurich.ibm.com}
\affiliation{IBM Research Europe - Zurich, 8803 R\"uschlikon, Switzerland}

\date{\today}

\maketitle

\textbf{Light-matter interaction enables engineering of non-equilibrium quantum systems. In condensed matter, spatially and temporally cyclic Hamiltonians are expected to generate energy-periodic Floquet states, with properties inaccessible at thermal equilibrium. A recent work explored the tunnelling conductance of a planar Josephson junction under microwave irradiation, and interpreted replicas of conductance features as evidence of steady Floquet-Andreev states. Here we realise a similar device in a hybrid superconducting-semiconducting heterostructure, which utilises a tunnelling probe with gate-tunable transparency and allows simultaneous measurements of Andreev spectrum and current-phase relation of the planar Josephson junction. We show that, in our devices, spectral replicas in sub-gap conductance emerging under microwave irradiation are caused by photon assisted tunnelling of electrons into Andreev states. The current-phase relation under microwave irradiation is also explained by the interaction of Andreev states with microwave photons, without the need to invoke Floquet states. The techniques outlined in this study establish a baseline to distinguish photon assisted tunnelling from Floquet-Andreev states in mesoscopic devices, a crucial development towards understanding light-matter coupling in hybrid nanostructures.}

%%% INTRODUCTION
Hybrid Josephson junctions (JJs) consist of a normal material (N) confined between two superconductors (S), where electron-hole reflection at the S-N interfaces leads to Andreev bound states (ABSs) at energies below the superconducting gap~\cite{Beenakker1991,Zazunov2003}. Microwave irradiation of SNS junctions has been shown to help elucidate fundamental properties of spin-orbit interaction and superconductivity~\cite{Fausti2011,Mitrano2016,VanWoerkom2017,Tosi2019}, drive coherent transitions between ABSs~\cite{Janvier2015,Larsen2015,deLange2015,Casparis2018,Wang2019,PitaVidal2020,Hays2021,Canadas2022}, and may facilitate the realisation of novel quantum states induced by light-matter interaction~\cite{Jiang2011,Bauer2019,Liu2019,Peng2021,Ji2022}.

An attempt to realise light-matter coupling was recently pursued using an aluminium/graphene SNS junction under microwave (MW) irradiation~\cite{Park2022}. Andreev bound states with energy $\Ea$ were measured with tunnelling spectroscopy via a superconducting lead, as shown by the density of states (DOS) schematic in the top panel of Fig.~\ref{fig1}(a). A current (yellow) flowed between a superconducting probe and an SNS junction, when occupied (red) and unoccupied (grey) states were aligned in energy by a source-drain bias $\Vsd$. Without microwave irradiation, this gave a differential conductance qualitatively similar to the blue curve in Fig.~\ref{fig1}(b), with peaks corresponding to ABSs at $\Vsd=\pm (\Delta+\Ea)/\mathrm{e}$, where $\Delta$ is the superconducting gap and $\mathrm{e}$ is the elementary charge. Under microwave irradiation with frequency $f$, replicas of ABSs and the superconducting gap edge, separated in voltage bias by $hf/\mathrm{e}$, were observed in the conductance spectrum, with $h$ being the Planck constant. These were interpreted in Ref.~\cite{Park2022} as signatures of steady Floquet-Andreev states (FASs) in the junction at energies $\Ea\pm nhf$, as schematically shown in the middle panel of Fig.~\ref{fig1}(a). Alternatively, photon assisted tunnelling (PAT)~\cite{Tien1963,Danchi1982,Platero2004,Roychowdhury2015,Kot2020,VanZanten2020,Peters2020,Carrad2022} can promote tunnelling across the barrier by absorption or emission of photons (green) with energy $hf$. The bottom panel of Fig.~\ref{fig1}(a) depicts an example of electron tunnelling into an ABS assisted by absorption of a photon. Both FASs and PAT give, at least qualitatively, conductance curves as shown in the bottom panel of Fig.~\ref{fig1}(b), with peaks at $\Vsd=(\Delta+\Ea\pm nhf)/\mathrm{e}$, where $n$ is an integer. Distinguishing the generation of FASs from PAT, which was not considered in Ref.~\cite{Park2022}, is crucial for the realisation of light-matter band engineering in nanoscale hybrid devices. Here we investigate conductance replicas emerging under microwave irradiation in the tunnelling spectrum of hybrid Josephson junctions in InAs/Al heterostructures. We perform multiple experimental tests to distinguish PAT from FASs and establish that, in our devices, PAT largely dominates the response to microwave irradiation. In the light of our results, conductance replicas which obey a sum rule are compatible with both FASs and PAT, and the tests we present here will likely provide a conclusive answer on their nature in future experiments. Our techniques are applicable to other devices and will guide towards the establishing of FASs in hybrid nanostructures.

\section{Results}
%%% FIG 1 - GATE-TUNABLE TUNNELLING BARRIER
\begin{figure*}
	\includegraphics[width=\textwidth]{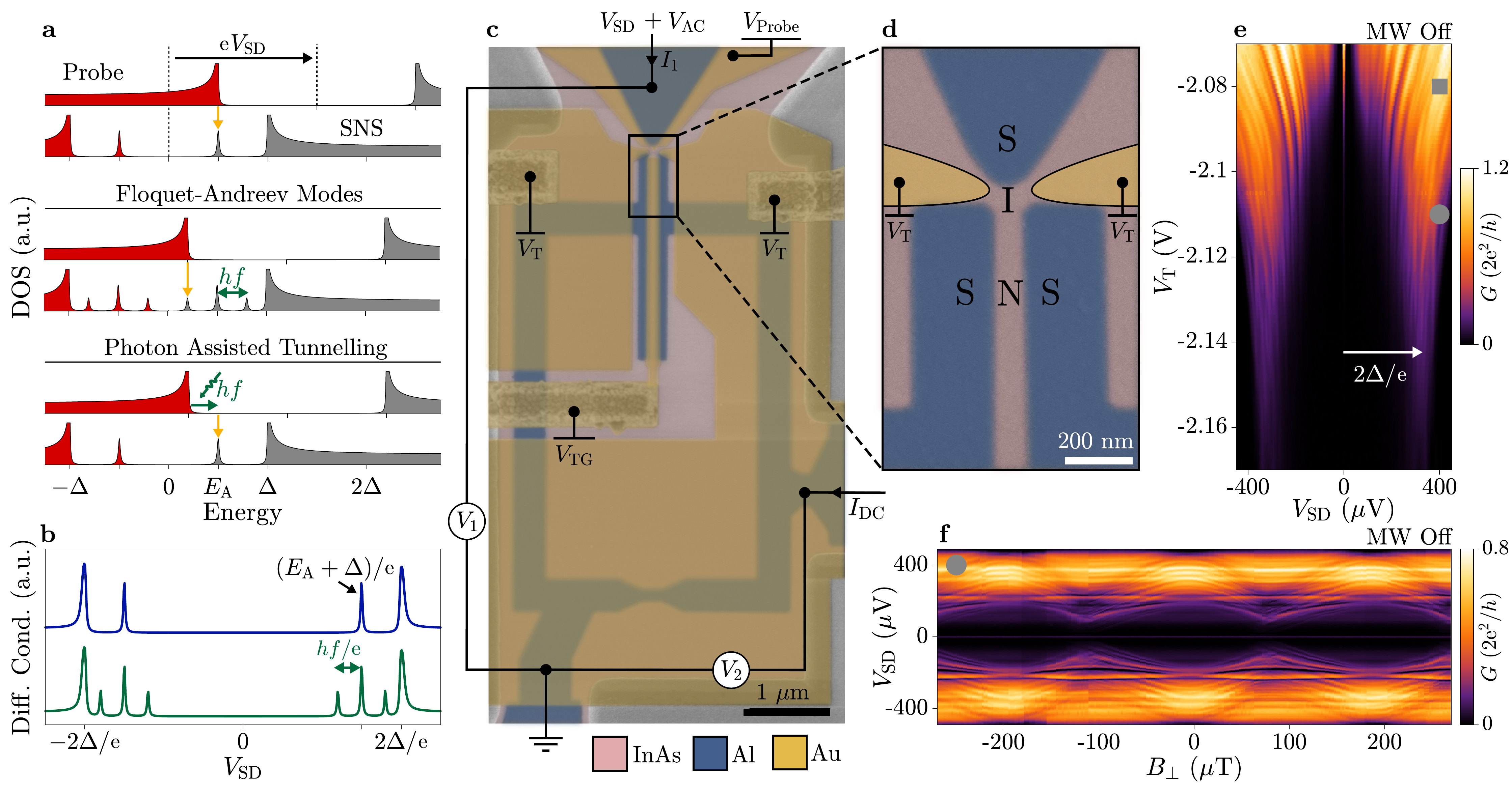}
	\caption{\textbf{Device under study and tunnelling spectroscopy of sub-gap states.} \textbf{(a)} Schematic representation of density of states (DOS) and tunnelling spectroscopy into a superconducting-semiconducting-superconducting (SNS) junction using a superconducting probe (top). Andreev bound states are present in the DOS of the SNS junction at energies $\pm\Ea$. A current (yellow) flows when the source-drain voltage $\Vsd$ aligns occupied (red) to unoccupied (grey) states. In a Floquet-Andreev scenario (middle), replicas of Andreev peaks shifted by the photon energy $hf$ emerge in the DOS of the SNS junctions, giving rise to additional tunnelling resonances. In a photon assisted tunnelling scenario (bottom), absorption of a photon (green) induces tunnelling into an ABSs for $\Vsd=\Ea-hf$. Floquet-Andreev modes are represented as replicas of Andreev peaks shifted by $hf$. \textbf{(b)} Schematic representation of tunnelling conductance measured in the absence (blue) and presence (green) of microwave irradiation. \textbf{(c)} False-coloured electron micrograph of a device identical to that under study, composed of InAs (pink) and Al (blue) and controlled via electrostatic gates (yellow). \textbf{(d)} Zoom-in of the tunnelling junction before gate deposition. The gates controlling the tunnelling barrier transparency are drawn in yellow. \textbf{(e)} Differential conductance $G$ of the tunnelling probe as a function of bias $\Vsd$ and gate voltage $\Vt$. The transport gap is indicated as $2\Delta/\mathrm{e}$. \textbf{(f)} Tunnelling spectroscopy of sub-gap states at $\Vt=-2.11~\mathrm{V}$, as a function of perpendicular magnetic field $\Bperp$.}
	\label{fig1}
\end{figure*}

Figure~\ref{fig1}(c) shows a false-coloured micrograph of the device presented in this Article. The device consisted of a planar SQUID fabricated in a heterostructure of InAs (pink) and epitaxial Al (blue)~\cite{Shabani2016,Fornieri2019,Nichele2020}, covered by a thin $\mathrm{HfO_2}$ insulating layer and with patterned Au gate electrodes (yellow). The superconducting loop contained a planar Al/InAs/Al junction and an Al constriction, all defined in the epitaxial Al. This Al constriction was designed to limit the switching current of the metal arm, while being significantly larger than that of the SNS junction. This configuration allowed for a stable phase drop across the SNS of $\varphi=2\pi(\mathit{\Phi}/\mathit{\Phi}_0)$, where $\mathit{\Phi}$ is the flux threading the SQUID and $\mathit{\Phi}_0=h/2\mathrm{e}$ is the superconducting flux quantum. The Al loop was connected to two low-impedance superconducting leads, which allowed switching current measurements. A gate-tunable superconducting tunnelling probe was integrated close to the SNS junction, allowing for spectroscopy into the normal region. Two gates controlled the transparency of the tunnelling probe by the gate voltage $\Vt$. The SNS junction was controlled by a top gate, which was set to $\Vtg=-0.8~\mathrm{V}$ for the results shown in the Main Text, unless otherwise stated. An additional gate was kept to $\Vprobe=0$ for the whole experiment. A zoom-in close to the tunnelling probe, obtained prior to gate deposition, is shown in Fig.~\ref{fig1}(d), where tunnelling gates are shown schematically. Microwave signals were applied via an attenuated coaxial line terminated in an antenna configuration and placed approximately $1~\mathrm{cm}$ away from the chip surface. Tunnelling conductance measurements were performed by lock-in techniques. A schematic of the measurement configuration is depicted in Fig.~\ref{fig1}(c). Tunnelling spectroscopy required sourcing a voltage bias $\Vsd+\Vac$ on the tunnelling probe lead and measuring the resulting AC current $\Ione$ and AC voltage $\Vone$. The current-phase relation (CPR) was measured by applying a DC current $\Idc$ through the loop and measuring the resulting DC voltage $\Vtwo$, which gave the loop switching current. The CPR of the SNS junction was obtained by subtracting the known switching current of the Al constriction from that of the loop. Measurements were performed in a dilution refrigerator at a mixing chamber base temperature of $7~\mathrm{mK}$. Further details on materials, fabrication and measurement techniques are reported in the Methods section.

\begin{figure*}
	\includegraphics[width=\textwidth]{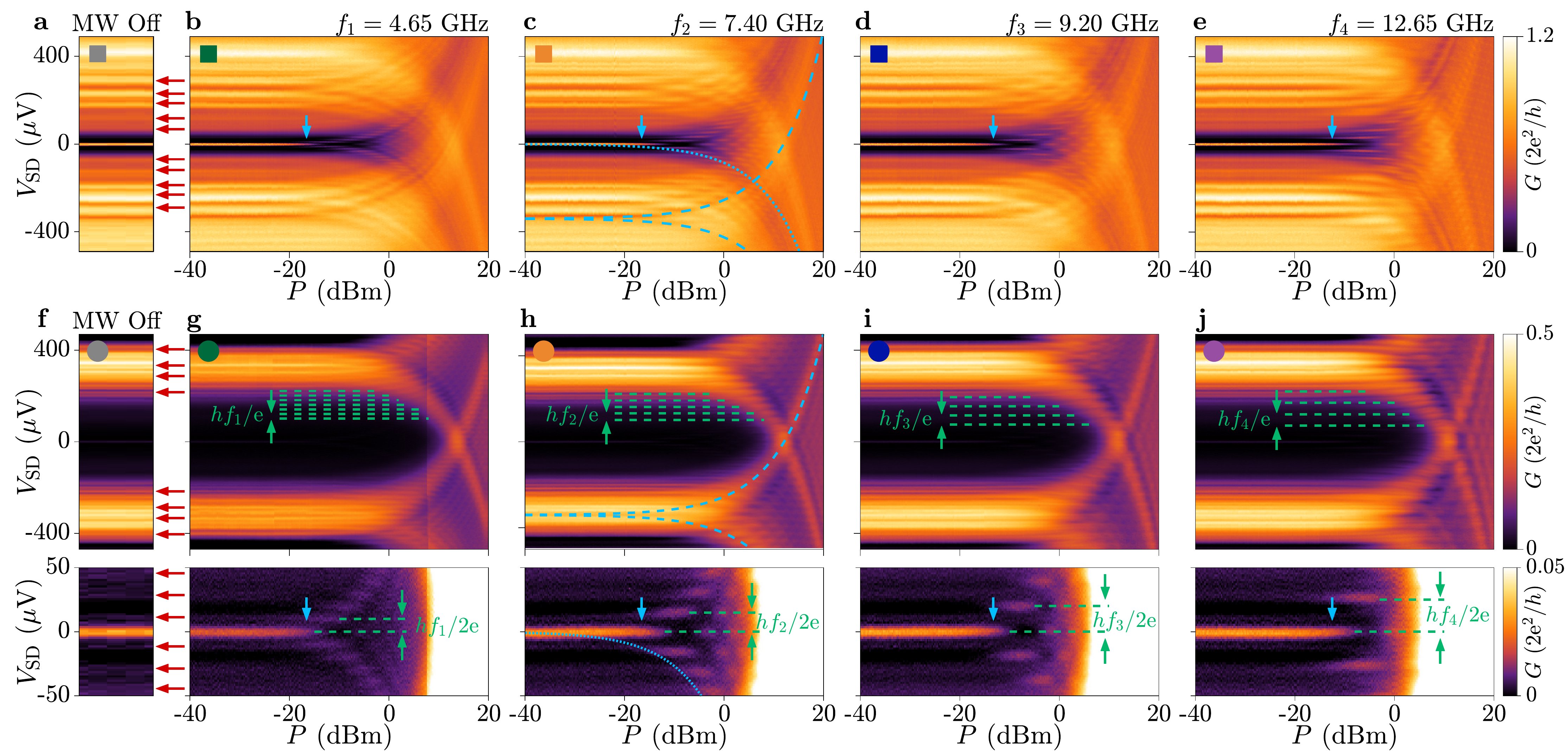}
	\caption{\textbf{Tunnelling conductance under microwave irradiation.} \textbf{(a)} Conductance at $\Vt=-2.08~\mathrm{V}$ with no microwave signal applied. Red arrows indicate sample-specific features present without irradiation. \textbf{(b-e)} Conductance at $\Vt=-2.08~\mathrm{V}$ for several irradiation frequencies as a function of microwave source power $P$ and $\Vsd$. The onset of splitting in conductance features is indicated by blue arrows. Blue dotted lines indicate the power dependence of split-conductance features. \textbf{(f)} As (a) but for $\Vt=-2.11~\mathrm{V}$. \textbf{(g-j)} As (b-e) but for $\Vt=-2.11~\mathrm{V}$ (top), with zoom-ins close to zero bias to highlight remnant supercurrent (bottom). Periodic replication of conductance features is indicated by green dashed lines.}
	\label{fig2}
\end{figure*}

% No RF conductance
Figure~\ref{fig1}(e) shows the differential conductance ${G\equiv\Ione/\Vone}$ of the tunnelling probe as a function of $\Vt$ and $\Vsd$, as the gate-tunable probe transitioned from the open to the tunnelling regime (top and bottom part of Fig.~\ref{fig1}(e), respectively). The open regime was characterised by a zero-bias conductance peak, which represents a supercurrent flowing through the tunnelling probe, and several finite bias features, which indicate multiple Andreev reflections. The tunnelling regime displayed pronounced features at a voltage $2\Delta/\mathrm{e}=380~\mathrm{\mu V}$ (white arrow), consistent with the superconducting gap $\Delta=190~\mathrm{\mu eV}$ of Al~\cite{Kjaergaard2017}. Figure~\ref{fig1}(f) shows $G$ at low barrier transparency ($\Vt=-2.11~\mathrm{V}$), as a function of perpendicular magnetic field $\Bperp$ and voltage bias $\Vsd$. Several finite bias conductance peaks are evident in Fig.~\ref{fig1}(f). In addition to some highly transmissive ABSs present within the superconducting gap of the SNS junction, additional features may result from multiple Andreev reflections, disorder in the tunnelling barrier and sub-gap states in the DOS of the superconducting probe~\cite{Su2018}.

\begin{figure}
	\includegraphics[width=\columnwidth]{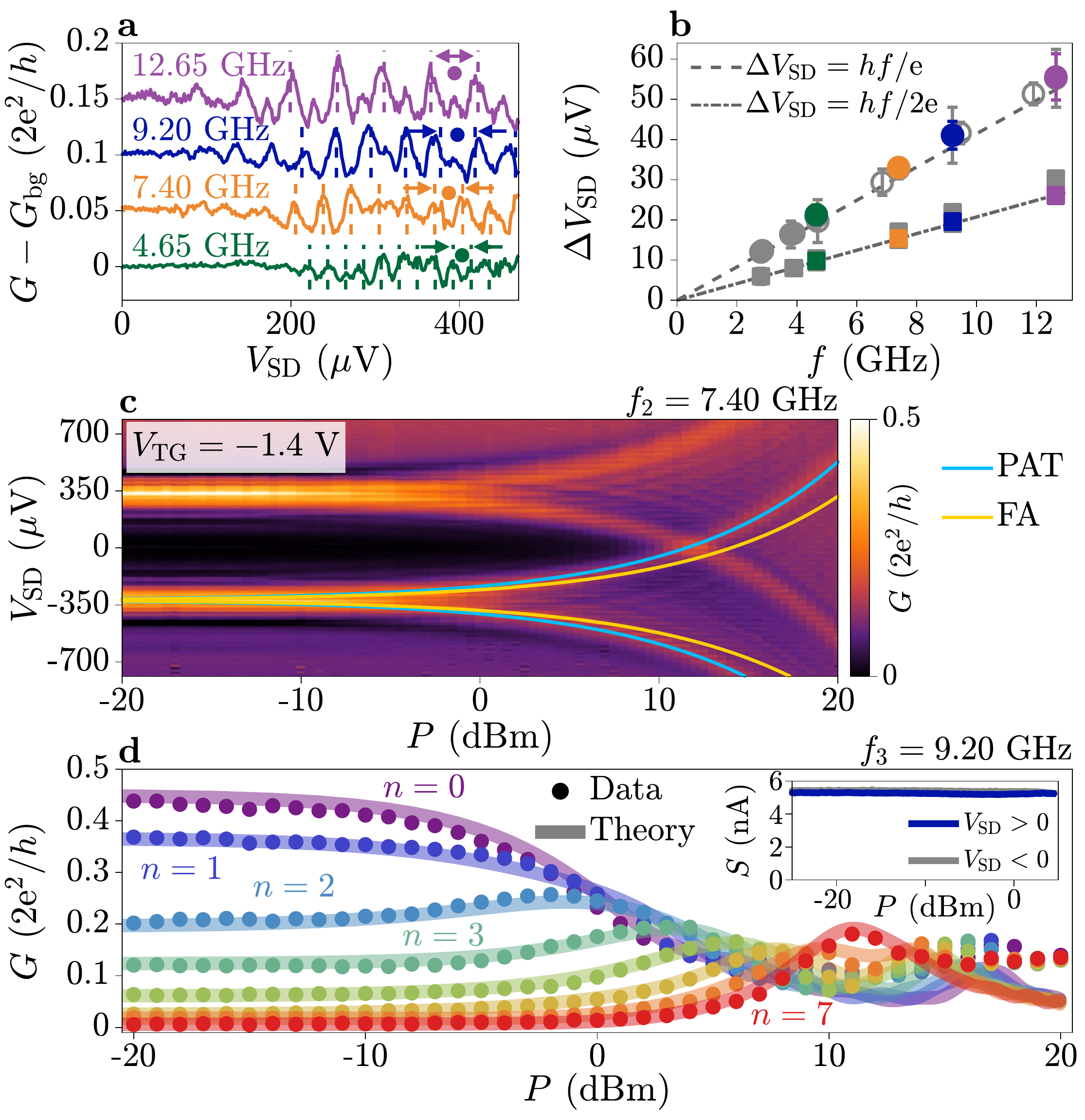}
	\caption{\textbf{Frequency and power dependence of conductance replicas.} \textbf{(a)} Linecuts of conductance from Fig.~\ref{fig2}, after subtraction of a slowly varying background. Traces are successively offset by $0.05~G_{0}$. Dashed lines mark the expected peak positions. Linecuts are taken at $P=1.5,~4.5,~4,~\mathrm{and}~4.5~\mathrm{dBm}$, respectively. \textbf{(b)} Spacing of conductance replicas measured at finite (circles) and close to zero (squares) bias. Dashed and dotted lines represent the equations $\Delta\Vsd=hf/\mathrm{e}$ and $\Delta\Vsd=hf/2\mathrm{e}$, respectively. Filled and empty grey markers refer to additional data collected on the same device and on a second device, respectively (see Supplementary Figs.~9-16). \textbf{(c)} Conductance at a top-gate voltage $\Vtg=-1.4~\mathrm{V}$, as a function of microwave source power for irradiation frequency $f_{2}=7.20~\mathrm{GHz}$. Blue lines indicate a microwave coupling strength of $\alpha_{0}=3.0$, identical to Figs.~\ref{fig2}(c,~h). Yellow lines indicate a coupling strength $\alpha_{0}=2.25$, reduced by $25\%$ with respect to Figs.~\ref{fig2}(c,~h). Blue (yellow) lines show the expectation for photon assisted tunnelling (Floquet-Andreev states). \textbf{(d)} Conductance of the first seven replicas in Fig.~\ref{fig2}(i), taken at constant bias $\Vsd$ (circles), alongside the simulated conductance from a photon assisted tunnelling model (lines, see Supplmentary Figs.~3-5). Inset shows the sum $S$ of conductance features in Fig.~\ref{fig2}(i) over positive (blue) and negative (grey) bias. Data is shown for the range of powers where all conductance replicas are within the measured bias range.}
	\label{fig3}
\end{figure}

%%% MICROWAVE RESPONSE OF JUNCTION
%% Distinguishing features from microwave influence
The effect of microwave irradiation on the tunnelling conductance $G$ for $\Bperp=0$ is summarised in Fig.~\ref{fig2}. The top row shows $G$ measured with the tunnelling probe in the open regime ($\Vt=-2.08~\mathrm{V}$), where the supercurrent flowing through the tunnelling probe is clearly visible. Figure~\ref{fig2}(a) shows $G$ without microwave irradiation, while Figs.~\ref{fig2}(b-e) show $G$ as a function of microwave source power $P$ at frequencies ${f_{i}=\{4.65,~7.40,~9.20,~12.65\}~\mathrm{GHz}}$. Similar plots, obtained in a more closed regime ($\Vt=-2.11~\mathrm{V}$) are plotted in Figs.~\ref{fig2}(f-j). In these plots, a remnant of the supercurrent is still visible by saturating the colourscale and zooming in close to zero bias, as we show in the bottom row of Fig.~\ref{fig2}.
For each frequency, conductance features at both zero and finite bias split into replicas as $P$ increased. Crucially, all conductance features split at the same power (see blue arrows) and evolved in an identical fashion as a function of $P$ as $\Vsd=(hf/\mathrm{e})\alpha$, where $\alpha=\alpha_{0}\cdot10^{P/20}$. This was true for each frequency investigated. An exemplary fit to the data of Fig.~\ref{fig2}(c), which yields $\alpha_{0}=3.0\pm0.4$, is plotted as the dotted blue line. Spacing between conductance replicas is indicated by green dashed lines, and is measured as $hf_{i}/\mathrm{e}$ and $hf_{i}/2\mathrm{e}$ for finite and zero bias features, respectively. While performing this analysis, it is important to distinguish conductance peaks that exclusively appear under microwave irradiation, to those already present without irradiation and that are caused by sample-specific features such as multiple Andreev reflections or sub-gap states in the superconducting probe [see red arrows Figs.~\ref{fig2}(a) and (f)]. Selected linecuts of Figs.~\ref{fig2}(g-j) are presented in Fig.~\ref{fig3}(a), after subtraction of a slowly varying background, together with a periodic grid with spacing $hf_{i}/\mathrm{e}$. Figure~\ref{fig3}(b) summarises the spacing between conductance replicas as a function of microwave frequency, which also includes additional frequencies, another $\Vtg$ value and a second device [see Supplementary Figs.~9-16]. Conductance replicas at finite bias are indicated by circles and depend on frequency as $\Delta \Vsd =hf/\mathrm{e}$ (dashed line). Supercurrent replicas are indicated as squares and follow the dependence $\Delta \Vsd =hf/2\mathrm{e}$ (dashed-dotted line). 

An example of conductance replicas at $\Vtg=-1.4~\mathrm{V}$ is shown in Fig.~\ref{fig3}(c). Blue lines indicate a coupling strength $\alpha_{0}=3.0$, identical to that in Figs.~\ref{fig2}(c,~h). Decreasing the top-gate voltage from $\Vtg=-0.8~\mathrm{V}$ to $\Vtg=-1.4~\mathrm{V}$ is expected to reduce the Fermi velocity by $\sim25\%$. In a model of FASs, $\alpha$ is proportional to the Fermi velocity~\cite{Park2022}, and therefore, assuming that all other parameters in the system stay the same, $\alpha$ is predicted to decrease by the same factor in Fig.~\ref{fig3}(c) relative to Figs.~\ref{fig2}(c,~h) (yellow lines). However, there is no observed change in the microwave coupling strength as a function of $\Vtg$, consistent with conductance replicas induced by PAT in the tunnel barrier and incompatible with FASs generated in the SNS junction (see Supplementary Material for further details). 

Finally, we present the power dependence of conductance replicas shown in Fig.~\ref{fig2}. Linecuts of Fig.~\ref{fig2}(i) are shown in Fig.~\ref{fig3}(d), for $n=7$ replicas (coloured circles). Their power dependence is modelled by a theory for PAT (lines)~\cite{Tien1963,Platero2004}, in which the conductance scales as a squared Bessel function $G\propto J^{2}_{n}(\alpha)$. This is similar to the theory used in Ref.~\cite{Park2022}, which follows the same dependence and differs mainly in the definition of $\alpha$. The PAT model takes two input parameters: the low-power conductance [Fig.~\ref{fig2}(f)] and $\alpha_{0}$, which is calculated from the low-bias conductance replicas [see Supplementary Figs.~3-5]. With no free parameters, a good agreement with the data is obtained up to $P\approx10~\mathrm{dBm}$, corresponding to $\alpha\approx8$. This demonstrates that high bias conductance features are explained by PAT, with an identical coupling strength to the microwave field as those at low bias. The sum over conductance features $S$ is constant for the range of powers where conductance replicas remain within the measured range of bias $\Vsd$ [see inset of Fig.~\ref{fig3}(d)]. This is consistent with conservation of the tunnel current, and stems from the result that squared Bessel functions sum to unity. Hence, the sum rule argument cannot be used to distinguish between PAT and FA interpretations [see Supplementary Material for more details].

Measurements presented in Figs.~\ref{fig2} and~\ref{fig3} demonstrate that conductance replicas originating from the supercurrent and from finite bias features (superconducting gap edge and ABSs) have identical coupling strength to the applied microwave field, and that the coupling strength does not change with gate voltage. Conductance replicas in the supercurrent are readily interpreted as Shapiro steps~\cite{Tinkham2004}, which only occur by photon absorption or emission in the tunnel barrier. Furthermore, all conductance replicas can be described by a PAT model up to large irradiation powers and the coupling strength was shown to be independent of top-gate voltage $\Vtg$, incompatible with FASs. We therefore conclude that finite bias replicas of Figs.~\ref{fig2} and~\ref{fig3} originate from PAT of electrons through the tunnelling barrier, and are not a manifestation of replicas in the DOS of the SNS junction.

% ENERGY SCALING WITH FREQUENCY

%%% CPR
\begin{figure}
	\includegraphics[width=\columnwidth]{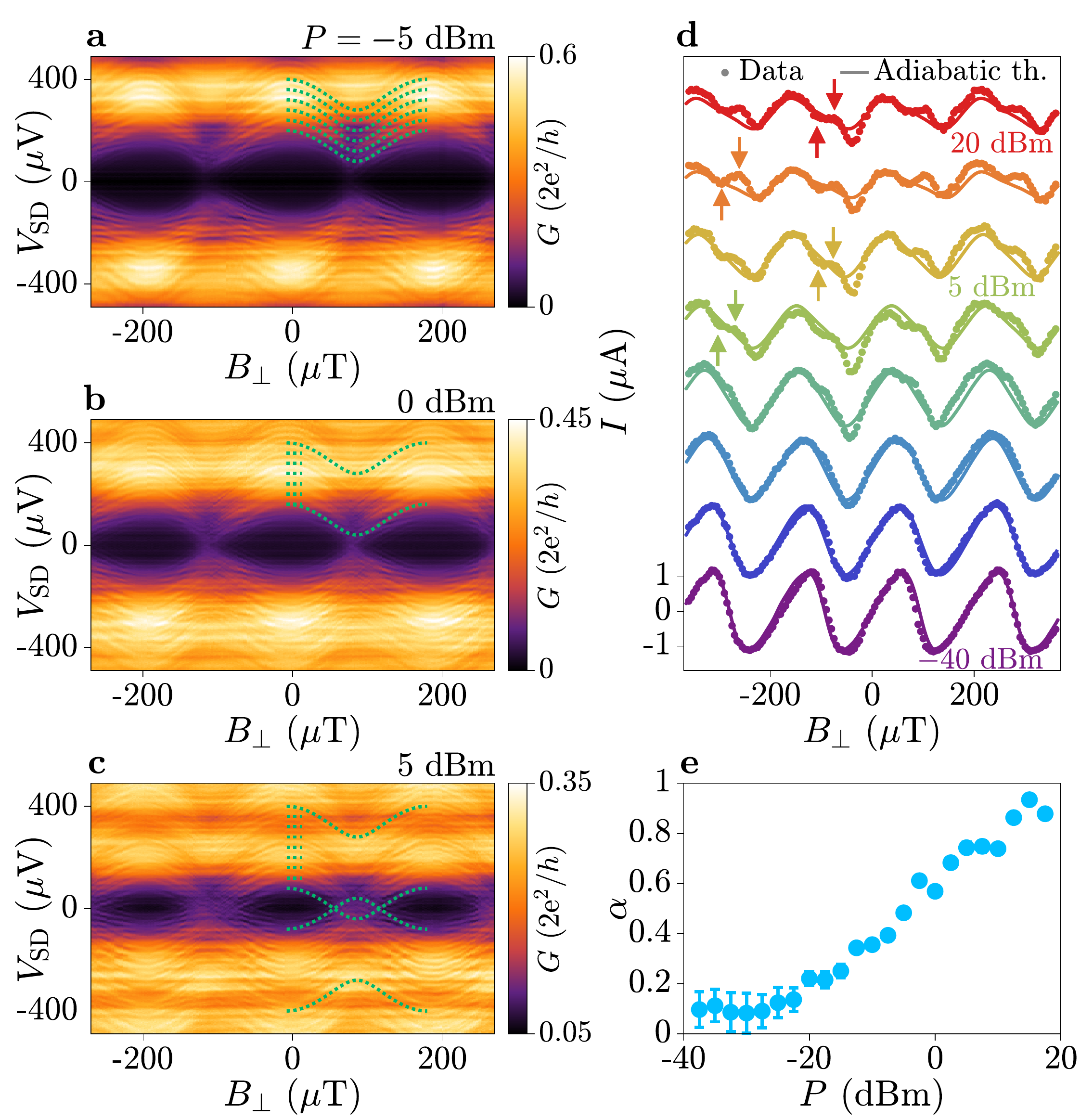}
	\caption{\textbf{Phase-dependent measurements under microwave irradiation.} \textbf{(a)} Tunnelling conductance as a function of perpendicular magnetic field $\Bperp$ for microwave irradiation with frequency $f=9.20~\mathrm{GHz}$ and power $P=-5~\dBm$. Green dashed lines mark bound state replicas with effective transmission $\bar{\tau}=0.84$. \textbf{(b, c)} Same as (a) for $P=0~\dBm$ and $P=5~\dBm$, respectively. \textbf{(d)} Current-phase relation as a function of microwave power (circles) fitted with an adiabatic theory (lines). Traces are successively offset by $2.05~\mathrm{\mu A}$. Deviations of the data from the adiabatic theory at high power are marked with arrows. \textbf{(e)} Microwave field strength $\alpha$ obtained from the adiabatic theory fit presented in (d).}
	\label{fig4}
\end{figure}

Similar to Ref.~\cite{Park2022}, we observed phase modulation of replicas originating from ABSs with energy ${\Ea=\Delta\sqrt{1-\bar{\tau}\sin^{2}(\varphi/2)}}$, where $\bar{\tau}$ is the effective junction transmission~\cite{Beenakker1991}. Figures~\ref{fig4}(a-c) show the tunnelling conductance as a function of perpendicular magnetic field $\Bperp$ measured under microwave irradiation of frequency $f=9.20~\mathrm{GHz}$ and applied power $P=-5$, $0$ and $5~\dBm$, respectively. Replicas of phase-dependent ABS features are indicated by green dashed lines, which describe ABSs with transmission $\bar{\tau}=0.84$. When $P$ was increased, more replicas appeared in the spectrum [Fig.~\ref{fig4}(b)] until replicas originating from positive and negative bias overlapped [Fig.~\ref{fig4}(c)]. No avoided crossing was observed for overlapping conductance features, in disagreement with FA predictions~\cite{Oka2009,Rudner2020,Liu2019}. 

After demonstrating that spectral replicas at high bias are caused by PAT in the tunnelling junction used to perform spectroscopy, we investigate how ABSs in the SNS junction couple to the applied electromagnetic field. These experiments probe the macroscopic superconducting state and do not rely on the microscopic processes taking place within the tunnelling probe, which was left floating. In particular, each occupied ABS in the SNS carries a supercurrent $I=-(2\mathrm{e}/\hbar) ( \partial E_{\mathrm{A}}(\varphi) / \partial \varphi)$. The total supercurrent flowing in the SNS is obtained by summing the contributions of each ABS~\cite{Nichele2020}. Figure~\ref{fig4}(d) shows the CPR of the SNS junction as a function of microwave power. For $P<-20~\mathrm{dBm}$, we observed a forward-skewed CPR, which indicates the presence of highly-transmissive ABSs~\cite{Beenakker1991,Spanton2017,Nichele2020}, consistent with the spectrum in Fig.~\ref{fig1}(f).

Increasing the applied microwave power, both the amplitude and skewness of the CPR decreased. This behaviour is described by an adiabatic theory of ABSs with a time-varying phase $\phi(t) = \varphi + 2\alpha\cos(2\mathrm{\pi} ft)$, where the electromagnetic field strength is $\alpha=\mathrm{e}V_{\mathrm{MW}}/hf$~\cite{Bergeret2011,Dou2021}. In this framework, the adiabatic current is given by ${\Iad=\Sigma_{n}\Jzero(2n\alpha)\In(\varphi)}$, where $\Jzero$ is a Bessel function of the first kind and $\In(\varphi)$ are the experimentally-determined harmonics of the CPR at equilibrium~\cite{Bergeret2011}. A fit with $\alpha$ as the sole free parameter describes the data well (solid lines), with fitted values of $\alpha$ shown in Fig.~\ref{fig4}(e). At $\alpha=0.6$ we extract $\alpha_{0}=0.3$, significantly smaller than $\alpha_{0}=2.6\pm0.1$ obtained for $f_{3}=9.20~\mathrm{GHz}$ from Fig.~\ref{fig2}(i) [see Supplementary Fig.~3]. This indicates that the coupling strength of the microwave field to the ABSs in the SNS junction, which is the parameter controlling the formation of FASs, is much smaller than that extracted from spectral replicas. This discrepancy is fully consistent with a PAT origin of spectral replicas, not linked to processes taking place in the SNS junction.

The adiabatic theory describes the data well up to an applied power of $P\sim5~\dBm$, corresponding to $\alpha\sim0.6$. For larger $P$, the adiabatic model still captures the CPR envelope, but does not account for dips in the CPR appearing at specific values of $\Bperp$ [arrows in Fig.~\ref{fig4}(d)]. Supercurrent dips are explained by non-equilibrium ABS occupation due to absorption or emission of microwave photons~\cite{Bergeret2011,Virtanen2010,Fuechsle2009}. Transitions occur when the photon energy $hf$, or integer multiples of it, matches the separation between two ABSs or between an ABS and the continuum. Once a transition occurs across the gap, the newly occupied ABS contributes to the total supercurrent with opposite sign with respect to the ground state, resulting in a dip in the CPR. Dips are resolved in the CPR for $\alpha>0.6$, consistent with a model of non-equilibrium ABS distribution in a JJ containing highly-transmissive modes [see Supplementary Fig.~19]. Therefore, the dominant effects on the CPR in our devices are an adiabatic modulation of the phase and a non-equilibrium distribution function of ABS occupation, up to $\alpha\gtrsim1$. It is possible that supercurrent signatures of FASs would emerge for larger microwave powers, but conductance features would be masked by the much stronger PAT effects.

\section{Discussion}
In summary, conductance replicas were realised in a hybrid Josephson junction with highly transmissive ABSs under microwave irradiation. They also obeyed a sum rule, consistent with both FAS and PAT interpretations. By performing additional tests, conductance replicas in our devices were shown to be inconsistent with FASs and instead caused by PAT, an effect not considered in Ref.~\cite{Park2022}. First, the power dependence of conductance replicas was identical to that of Shapiro steps in the tunnelling junction, whereas a difference is expected for FASs. Second, the coupling strength $\alpha$ associated with conductance replicas was significantly smaller than that associated with ABSs in the SNS junction and measured via switching currents, but should be equal in the case of FASs. Third, the coupling strength was independent on the Fermi velocity, inconsistent with the linear dependence predicted for FASs. Fourth, conductance replicas brought to zero energy crossed each other, while anti-crossing is expected for FASs. Complementary measurements of the current-phase relation of the Josephson junction are consistent with an interaction between ABSs and the microwave field mediated by the superconducting phase difference, without the need to invoke FASs. The weak coupling of the microwave field to ABSs is presumably due to the use of an off-chip microwave antenna, which predominantly interacts with the device via the large leads. Future work can engineer more efficient coupling schemes, for example by applying local microwave signals via gate electrodes~\cite{Tosi2019}, enabling stronger interaction with ABSs while limiting heating in the setup.

Our results show that caution should be used to attribute replicas in the tunnelling conductance to the presence of Floquet states in hybrid Josephson junctions. However, the techniques outlined here constitute a baseline to evaluate the effect of light-matter interaction in nanoscale devices, as they give distinct signatures for FASs and PAT, and can be applied in generic cases.

\section*{Acknowledgements}
We are grateful to C.~Bruder, W.~Riess and H.~Riel for helpful discussions. We thank the Cleanroom Operations Team of the Binnig and Rohrer Nanotechnology Center (BRNC) for their help and support. F.~N. acknowledges support from the European Research Council (grant number 804273) and the Swiss National Science Foundation (grant number 200021\_201082). W.~B. acknowledges support from the European Union's Horizon 2020 FET Open programme (grant number 964398) and from  the Deutsche Forschungsgemeinschaft (DFG; German Research Foundation) via the SFB 1432 (ID 425217212). J.~C.~C. thanks the Spanish Ministry of Science and Innovation (Grant No. PID2020-114880GB-I00) for financial support and the DFG and SFB 1432 for sponsoring his stay at the University of Konstanz as a Mercator Fellow.

\section*{Data Availability}
Data presented in this work will be available on Zenodo. The data that support the findings of this study are available upon reasonable request from the corresponding author.

\bibliography{Bibliography1}

\section*{Methods}
\subsection*{Materials and Fabrication}
The devices under study were fabricated from a heterostructure grown on an InP (001) substrate by molecular beam epitaxy techniques. The heterostructure consisted of a step-graded metamorphic InAlAs buffer and a $7~\mathrm{nm}$ thick InAs quantum well, confined by $\mathrm{In_{0.75}Ga_{0.25}As}$ barriers $13~\mathrm{nm}$ below the surface. A $10~\mathrm{nm}$ thick Al layer was deposited on top of the heterostructure, in the same chamber as the III-V growth without breaking vacuum. The peak mobility in a gated Hall bar was $18000~\mathrm{cm^{2}V^{-1}s^{-1}}$ for an electron density of $n=8\cdot10^{11}~\mathrm{cm^{-2}}$. This gave an electron mean free path of $l_{e}\gtrapprox270~\mathrm{nm}$, hence we expect all Josephson junctions measured here to be ballistic along the length $L$ of the junction.

Devices were fabricated by first isolating large mesa structures in the III-V material to prevent parallel conduction between devices. This was done by selectively removing the top Al layer with Al etchant Transene D, before etching $\sim350~\mathrm{nm}$ into the III-V heterostructure using a chemical wet etch ($220:55:3:3$ solution of $\mathrm{H_{2}O:C_{6}H_{8}O_{7}:H_{3}PO_{4}:H_{2}O_{2}}$). The planar SQUID device was then patterned on top of the mesa structure, by selective etching of the Al with Transene D at $50\mathrm{^{\circ}C}$ for $4~\mathrm{s}$. We deposited a dielectric by atomic layer deposition, consisting of a $3~\mathrm{nm}$ $\mathrm{Al_{2}O_{3}}$ layer below $15~\mathrm{nm}$ of $\mathrm{HfO_{2}}$, before evaporating metallic gate electrodes to control the exposed III-V regions. These were deposited in two steps: fine gate features above the planar SQUID were first defined with $5~\mathrm{nm}$ of Ti and $20~\mathrm{nm}$ of Au; these were contacted with $10~\mathrm{nm}$ of Ti and $400~\mathrm{nm}$ of Al, to connect the mesa structure to the bonding pads.

\subsection*{Measurement Techniques}
Measurements were performed in a dilution refrigerator with a base temperature of $7~\mathrm{mK}$. Conductance measurements were performed with standard lockin-amplifier techniques. An AC voltage $\Vac=3~\mathrm{\mu V}$ was applied to a contact at the superconducting probe with a frequency of $311~\mathrm{Hz}$. The current flowing through the probe to ground, $\Ione$, and the differential voltage across the tunnel barrier $\Vone$ were measured to give the conductance $G\equiv\Ione/\Vone$. The transmission from the probe to the SNS junction was tuned using the gates $\Vt$. In the tunnelling regime, where the conductance is lower than one conductance quantum $G_{0}=2\mathrm{e}^{2}/h$, the measured conductance is the convolution of the density of states in the probe and the junction: $G_{\mathrm{Meas.}}=G_{\mathrm{Probe}} \ast G_{\mathrm{SNS}}$. A constant bias offset of $43~\mathrm{\mu V}$ was subtracted from all datasets, due to a DC offset at the current-voltage (I-V) converter. The plotted bias voltage $\Vsd$ at the device was adjusted from the sourced value to account for the voltage dropping across a measured line resistance of $6.4~\mathrm{k\Omega}$. 
Current-biased measurements were performed on the same device. Both contacts at the superconducting probe were floated, such that no current flowed through the probe. A DC current $\Idc$ was applied symmetrically to the SQUID loop, such that the potential of the device was not raised with respect to the gate electrodes. A sawtooth current signal was applied from a waveform generator at a frequency of $133~\mathrm{Hz}$. The voltage drop across the SQUID loop $\Vtwo$ was measured with an oscilloscope. Once the voltage passed a threshold signifying a superconducting-resistive transition, the switching current was recorded. To account for stochastic phase escape behaviour, the average switching current was recorded over $36$ values. 
A high-frequency signal was applied to the device via an antenna $\sim 1~\mathrm{cm}$ away from the surface of the chip. The antenna, an exposed coaxial line, was attached to a microwave line with attenuation $47~\mathrm{dB}$. Powers $P$ refer to the power outputted at the signal generator.

\clearpage
\newpage

\onecolumngrid

\section*{Supplementary Material: Microwave-induced conductance replicas in hybrid Josephson junctions without Floquet-Andreev states}

\newcounter{myc} %define new counter
\newcounter{myc2} %define new counter
\renewcommand{\thefigure}{S.\arabic{myc}}
\renewcommand{\theequation}{S.\arabic{myc2}}
\renewcommand{\thesection}{\arabic{section}}

\section{Frequency Dependence of Conductance Response}
\setcounter{myc}{1}
\begin{figure}[h]
	\includegraphics[width=0.5\columnwidth]{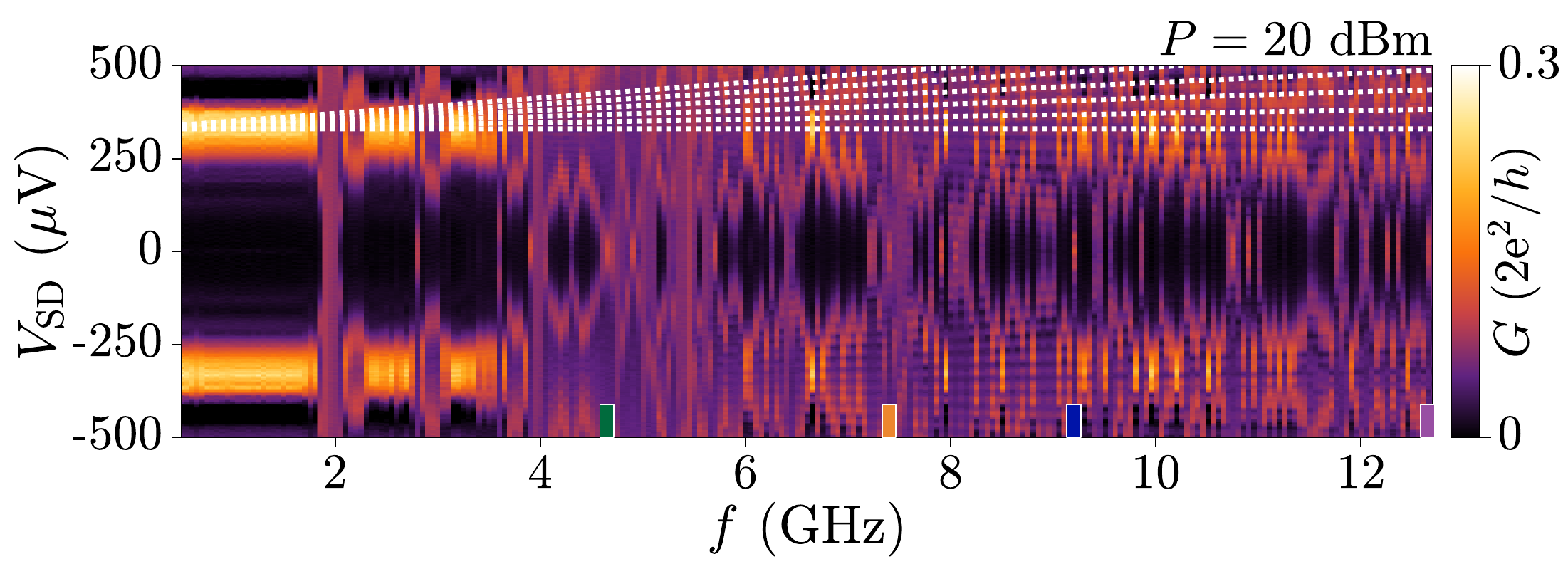}
	\caption{\textbf{Frequency dependence of conductance $G$ as a function of source-drain bias $\Vsd$, at a fixed microwave power $P=20~\mathrm{dBm}$.} Conductance replicas are schematically indicated by the dashed white lines, $\Delta\Vsd=nhf/\mathrm{e}$. Frequencies shown in Figs. 2 and 3 of the main text are indicated by coloured markers.}
	\label{Sfig1}
\end{figure}
The microwave response of the device was first investigated as a function of microwave (MW) irradiation with frequency $f$, in the low barrier transparency regime ($\Vt=-2.11~\mathrm{V}$). Figure~\ref{Sfig1} shows the differential conductance $G$ as a function of source-drain bias $\Vsd$ for increasing frequency from $500~\mathrm{MHz}$ to $12.7~\mathrm{GHz}$, with an applied power $P=20~\mathrm{dBm}$. The conductance was unaffected by the applied signal for frequencies up to $1.8~\mathrm{GHz}$. At frequencies $f>1.8~\mathrm{GHz}$, the conductance was altered by the applied microwaves and in some cases we observed a non-zero conductance at $\Vsd=0$. The conductance response to irradiation frequency was non-monotonic, suggesting that the coupling strength of the antenna to the device was frequency-dependent. This was due to the method of applying microwaves by an exposed antenna within the sample space. The results shown in Figs. 2 and 3 of the Main Text were measured at frequencies labelled by the coloured markers. These frequencies were chosen where the response of the conductance was strongest based on Fig.~\ref{Sfig1}, such that a full power dependence was possible.

Replication of conductance features, as those in the Main Text, were evident at both positive and negative bias. These followed a linear dependence on frequency, as highlighted by the white dashed lines at positive bias. Conductance replicas follow the relation $\Vsd=nhf/\mathrm{e}$, where $n$ is an integer. Hence, the separation of conductance replicas of $\Delta\Vsd=hf/\mathrm{e}$ was consistent across a wide frequency range.

\section{Spectroscopy at an In-Plane Magnetic Field}
\setcounter{myc}{2}
\begin{figure}[h]
	\includegraphics[width=0.5\columnwidth]{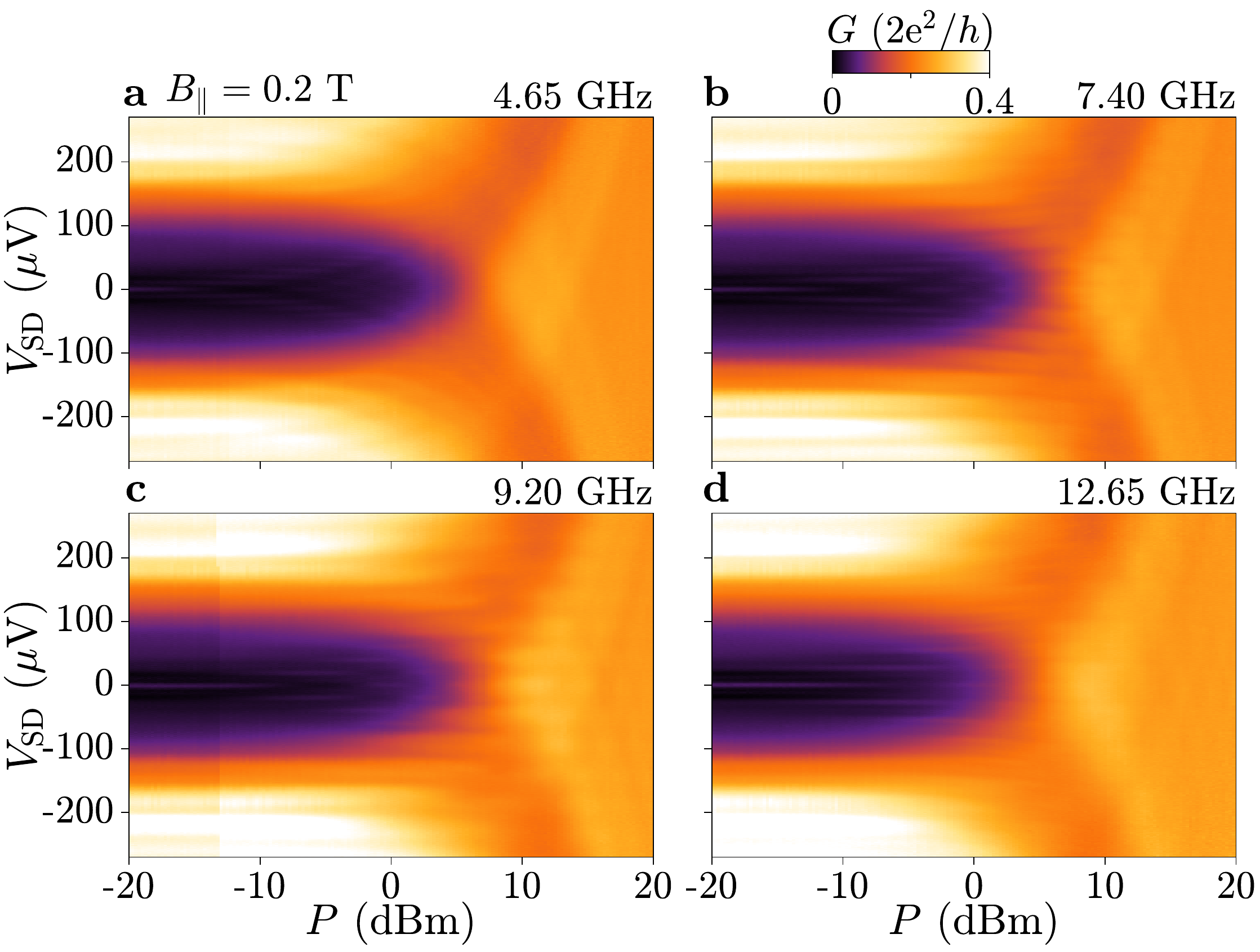}
	\caption{\textbf{Power dependence of conductance $G$ as a function of bias $\Vsd$, at an in-plane magnetic field $B_{\parallel}=0.2~\mathrm{T}$ for different irradiation frequencies ${f=\left\{4.65,~7.40,~9.20,~12.65\right\}~\mathrm{GHz}}$.}}
	\label{Sfig7}
\end{figure}

In the Main Text, results were shown for an in-plane magnetic field $B_{\parallel}=0$. Both the Josephson junction and the probe were in the superconducting state (S), meaning that tunnelling across the insulating barrier (I) corresponded to an SIS geometry [as shown in Fig.~1(a) of the Main Text]. At an in-plane magnetic field of $B_{\parallel}=0.2~\mathrm{T}$, superconductivity in the probe was suppressed such that there was a finite density of states within the superconducting gap of the probe. The differential conductance $G$ therefore showed features at bias values proportional to the density of states in the Josephson junction.

Figure~\ref{Sfig7} shows bias spectroscopy at an in-plane magnetic field of $B_{\parallel}=0.2~\mathrm{T}$ as a function of power $P$, for different frequencies $f$ of applied radiation. The device configuration was identical to the open regime outlined in the Main Text ($\Vt=-2.08~\mathrm{V}$). The conductance $G$ as a function of source-drain voltage $\Vsd$ shows a superconducting gap at low bias. Conductance values increased to a maximum close to $|\Vsd|=200~\mathrm{\mu V}$. A small conductance peak was visible at $\Vsd=0$, from a small residual supercurrent which flowed across the tunnel barrier. 

On increasing microwave power $P$, replicas in conductance features emerged at both high and low bias. High-bias conductance replicas had separation $\Delta\Vsd=hf/\mathrm{e}$, as seen in the Main Text. Furthermore, the power dependence was similar to that at $B_{\parallel}=0$. Conductance replicas were present under microwave irradiation when superconductivity was suppressed in the probe. This is consistent with photon assisted tunnelling (PAT) into Andreev bound states (ABSs) of charges in the probe at the Fermi energy. 

\section{Microwave Field Strength from Shapiro Steps}
The conductance of replicas appearing under microwave irradiation depends on the applied power $P$. We first considered the power dependence of Shapiro steps close to $\Vsd=0$ [see blue dotted lines in Figs.~2(c, h) of the Main Text]. Conductance peaks occurred when the source-drain bias $\Vsd$ was equal to the Josephson voltage $V_{\mathrm{J}}=nhf/2\mathrm{e}$, where $n$ is an integer denoting the order of the Shapiro step. The conductance of the $n^{\mathrm{th}}$ Shapiro step is proportional to the $n^{\mathrm{th}}$-order Bessel function of the first kind, $J_{n}(2\mathrm{e}\Vmw/hf)$, where $\Vmw$ is the amplitude of the oscillating voltage due to the applied microwave signal. This corresponds to the most likely number of photons absorbed in the system. This scales linearly with $n$, such that there is an almost exact correspondence between $\Vmw$ and the $\Vsd$ at which the highest conductance peak occurs. The applied microwave signal is given as a power $P$ in units of $\mathrm{dBm}$. We therefore express the oscillating voltage at the sample as $\Vmw=V_{0}\cdot10^{P/20}$, where $V_{0}$ contains the output voltage, device-antenna coupling and coaxial line attenuation of $47~\mathrm{dB}$. In dimensionless units, the coupling strength to the microwave field is therefore defined as $\alpha\equiv \mathrm{e}\Vmw/hf = (\mathrm{e}/hf)V_{0}\cdot10^{P/20}$.

\setcounter{myc}{3}
\begin{figure}[h]
	\includegraphics[width=\textwidth]{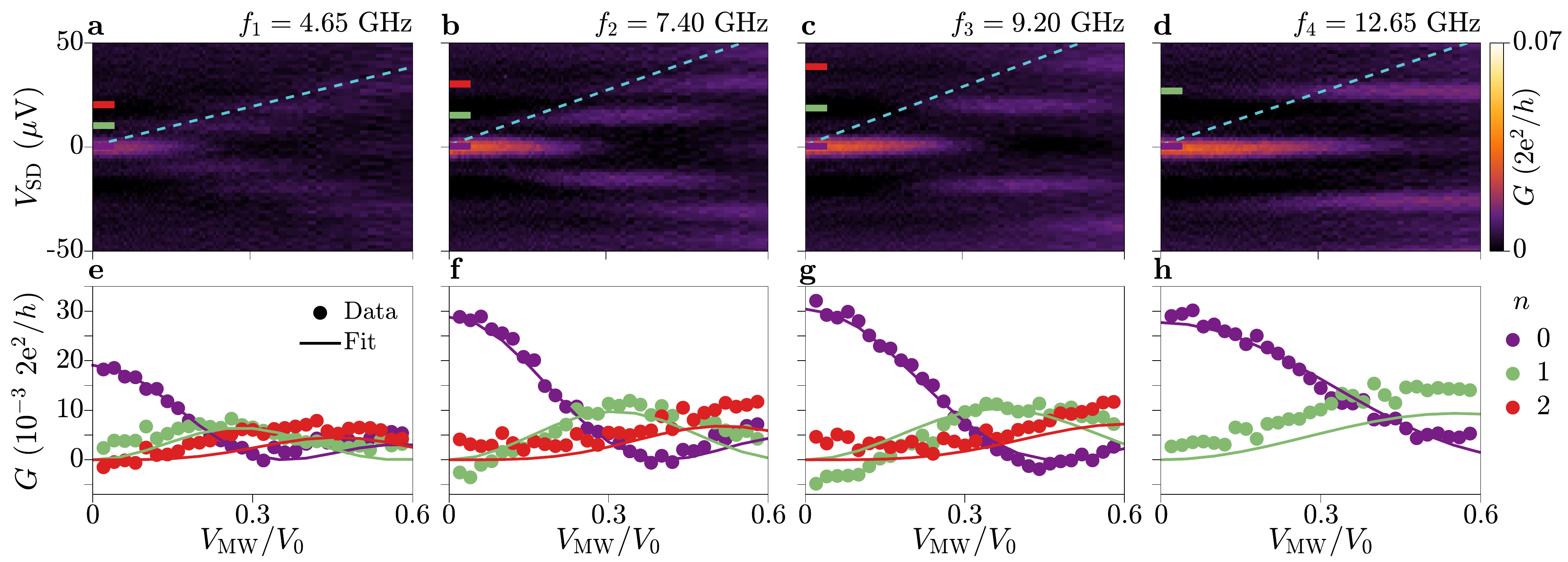}
	\caption{\textbf{Microwave field strength from Shapiro steps.} \textbf{(a-d)} Conductance of Figs.~2(g-j) of the Main Text, plotted as a function of microwave field strength $\Vmw/V_{0}=10^{P/20}$. Shapiro steps at $\Vsd=nhf/2\mathrm{e}$ are indicated by the coloured bars. Blue dashed lines indicate $V_{0}$, as calculated from a fit. \textbf{(e-h)} Linecuts of conductance in (a-d) at bias values corresponding to the $n=0,~1,~2$ order Shapiro steps (circles). Fit to the conductance (lines) to obtain the parameter $V_{0}$.}
	\label{Sfig15b}
\end{figure}

The Shapiro steps in the closed regime [Figs.~2(g-j) of the Main Text] are plotted in Fig.~\ref{Sfig15b}(a-d) as a function of microwave field strength $\Vmw/V_{0}$. The emergence of the $n^{\mathrm{th}}$ Shapiro step scales linearly with $\Vmw$ as indicated by the blue dashed lines, the gradient of which is given by $V_{0}$. Linecuts of the zeroth, first and second order Shapiro steps are plotted in Figs.~\ref{Sfig15b}(e-h) as the purple, green and red circles, respectively. The plotted data is sampled from the raw data at intervals $\Delta\Vmw/V_{0}=0.02$, to have a regular separation of datapoints. The conductance of the $n^{\mathrm{th}}$ Shapiro step, $G_{n}(\Vmw)\equiv G(\Vsd=nhf/2\mathrm{e},~\Vmw)$, is fitted with a squared Bessel function of the form~\cite{Tien1963,Tinkham2004,Platero2004,Peters2020,Kot2020}

\setcounter{myc2}{1}
\begin{equation}
	G_{n}(\Vmw) = G_{n}(\Vmw=0)\left[J_{n}\left(\frac{2\mathrm{e}}{hf}\Vmw\right)\right]^{2} = G_{n}(\Vmw=0)\left[J_{n}\left(\frac{2\mathrm{e}}{hf}V_{0}\cdot10^{P/20}\right)\right]^{2},
\end{equation}
with $G_{n}(\Vmw=0)$ the conductance at bias $\Vsd=nhf/2\mathrm{e}$ with no microwaves applied. The fit with the free parameter $V_{0}$ returns $V_{0} = \left\{64,~91,~96,~87\right\}~\mathrm{\mu V}$ and is plotted as the lines in Fig.~\ref{Sfig15b}(e-h) for frequencies $f_{1}$ to $f_{4}$, respectively. The corresponding dimensionless microwave field strengths are $\alpha_{0}=\{3.3,~3.0,~2.5,~1.7\}$.

\section{Modelling of Photon Assisted Tunnelling Data}
\setcounter{myc}{4}
\begin{figure}[h]
	\includegraphics[width=\textwidth]{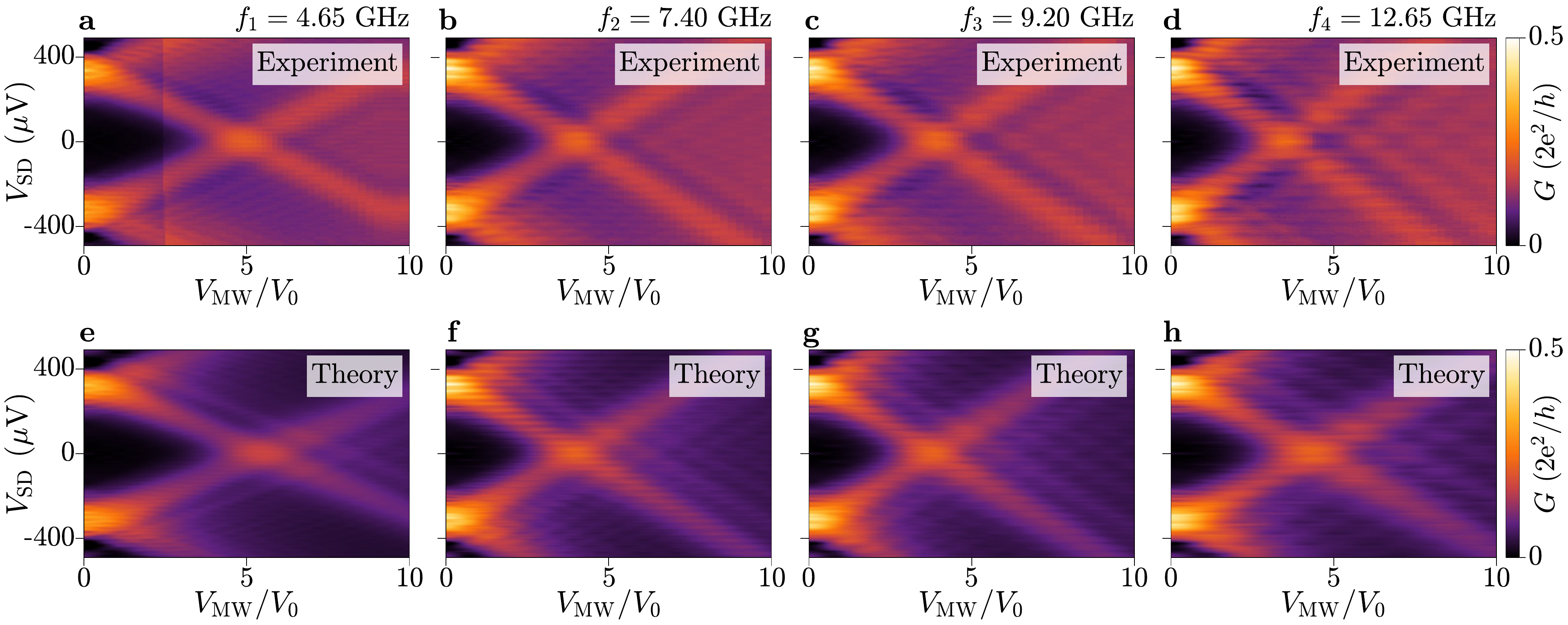}
	\caption{\textbf{Experimental and simulated conductance replicas as a function of microwave field strength.} \textbf{a-d} Differential conductance of Figs.~2(g-j) of the Main Text, plotted as a function of microwave field strength $\Vmw/V_{0}=10^{P/20}$. \textbf{(e-h)} Simulated conductance features as a function of microwave field strength, using the coupling parameters $V_{0}$ obtained in Fig.~\ref{Sfig15b} and the measured conductance in the absence of microwave irradiation.}
	\label{Sfig16}
\end{figure}

Figures~\ref{Sfig16}(a-d) show the conductance maps of Figs.~2(g-j) of the Main Text plotted as a function of microwave field strength $\Vmw/V_{0}$. Conductance replicas emerge linearly with increasing microwave field strength. The experimental data is simulated using a model for photon assisted tunnelling, based on the coupling parameters $V_{0}$ obtained from the Shapiro steps [see Fig.~\ref{Sfig15b}]. The $n^{\mathrm{th}}$-order conductance replicas are expected to scale as a squared Bessel function~\cite{Tien1963,Tinkham2004,Platero2004,Peters2020,Kot2020}:

\setcounter{myc2}{2}
\begin{equation}
	G\left(\Vmw,~\Vsd+n\frac{hf}{\mathrm{e}}\right) = G\left(\Vmw=0,~\Vsd+n\frac{hf}{\mathrm{e}}\right)\left[J_{n}\left(\frac{\mathrm{e}\Vmw}{hf}\right)\right]^{2}.
	\label{eqS2}
\end{equation}
Using the experimentally measured conductance with no applied microwaves, $G(\Vmw=0,~\Vsd)$, the conductance at each $\Vmw$ was calculated by summing the contributions from  $N$ replicas:

\setcounter{myc2}{3}
\begin{equation}
	G(\Vmw,~\Vsd) = \sum_{n=-N}^{N}G\left(\Vmw=0,~\Vsd+n\frac{hf}{\mathrm{e}}\right),
	\label{eqS3}
\end{equation}
where $N=(1~\mathrm{mV})\cdot\mathrm{e}/hf$ was chosen to consider conductance replicas emerging across the full range of measured source-drain bias. The simulated conductance is plotted in Figs.~\ref{Sfig16}(e-h) as a function of microwave field strength $\Vmw/V_{0}$, using the values of $V_{0}$ obtained from the Shapiro steps at each frequency. The replication of conductance features is well described by the simulation, up to the highest measured microwave fields, in terms of the number of replicas, their dependence of microwave field strength and the absolute value of their conductance. Some discrepancy at large $\Vmw/V_{0}$ can be attributed to a background conductance in the measurement data, potentially due to device heating which is not accounted for in simulation.

\setcounter{myc}{5}
\begin{figure}[h]
	\includegraphics[width=\textwidth]{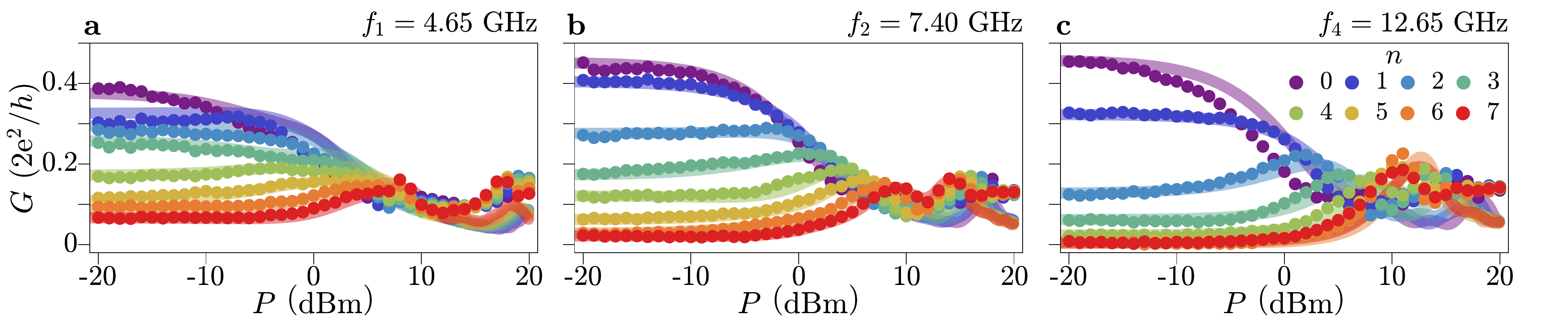}
	\caption{\textbf{Power dependence of conductance replicas.} Conductance of the first seven replica peaks in Fig.~2(g,~h,~j) of the Main Text (circles), respectively. Colours denote the order $n$ of the replica. Plotted alongside simulated conductance from Fig.~\ref{Sfig16} (lines), as a function of applied microwave power $P$.}
	\label{Sfig17}
\end{figure}

Figure~\ref{Sfig17} shows the conductance of replica peaks at fixed bias $\Vsd$ as a function of applied power $P$, up to the seventh replica [circles, replica number indicated by the colour]. Data is plotted for frequencies $f_{1}=4.65~\mathrm{GHz}$, $f_{2}=7.40~\mathrm{GHz}$ and $f_{4}=12.65~\mathrm{GHz}$, since the equivalent data for $f_{3}=9.20~\mathrm{GHz}$ is plotted in Fig.~3(c) of the Main Text. The simulated conductance at the same bias is plotted as the shaded lines, and matches the data for low and intermediate powers $P\lesssim10~\mathrm{dBm}$. Data is plotted for $P\gtrsim-20~\mathrm{dBm}$ to better highlight the power dependence, since only small changes in conductance were observed in the range $-40~\mathrm{dBm}<P\lesssim-20~\mathrm{dBm}$.

\section{Removal of Background Conductance}
\setcounter{myc}{6}
\begin{figure*}[h]
	\includegraphics[width=\textwidth]{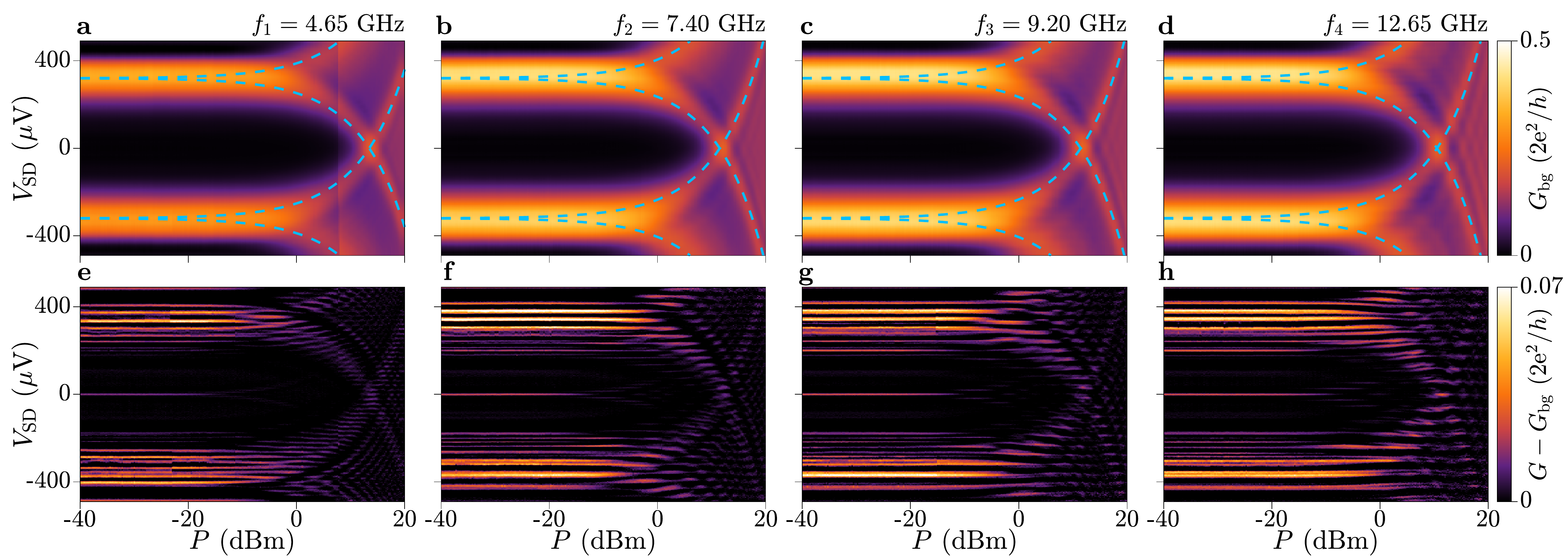}
	\caption{\textbf{Bias spectroscopy with the removal of a slowly-varying background} \textbf{(a-d)} Power dependence of Fig. 2 in the Main Text at different frequencies, averaged at each power across a bias voltage window $V_{\mathrm{window}}=70~\mathrm{\mu V}$. Dashed lines indicate the power dependence of high conductance features. \textbf{(e-h)} Power dependence of Fig. 2 in Main Text, with the averaged background removed $G-\Gbg$. Linecuts in Fig. 3(a) of the Main Text are taken at powers $P=\left\{1.5,~4.5,~4,~4.5\right\}~\mathrm{dBm}$, respectively.}
	\label{Sfig8}
\end{figure*}
The conductance maps in Fig.~2 of the Main Text show the complete response of the system to a microwave drive of increasing power. Figures~\ref{Sfig8}(a-d) show a slowly-varying background conductance $\Gbg$, obtained by averaging the conductance trace at each power $P$ over a bias window of $70~\mathrm{\mu V}$. Dashed lines show the dependence of high conductance features on power $P$, with the relation $\Vmw=V_{0}\cdot10^{P/20}$ for values of $V_{0}$ calculated from the Shapiro steps [see Fig.~\ref{Sfig15b}].

Conductance replicas were isolated by subtracting the slowly-varying background, $G-\Gbg$ [see Figs.~\ref{Sfig8}(e-h)]. The linecuts in Fig.~3(a) of the Main Text were taken at powers $P=1.5~\mathrm{dBm},~4.5~\mathrm{dBm},~4~\mathrm{dBm}~\mathrm{and}~4.5~\mathrm{dBm}$ from Figs.~\ref{Sfig8}(e-h) respectively, such that multiple conductance replicas were visible. The separation between conductance features $\Delta\Vsd$ shown in Fig.~3(b) of the Main Text was calculated by taking the average of conductance peak separation across the full power range displayed in Figs.~\ref{Sfig8}(e-h).

The background conductance contained features from the complex ABS spectrum at low power. Multiple high-conductance lines were visible, in both the background conductance and the difference, due to replication of different features in the low-power conductance map. Such conductance features could include sub-gap ABSs at $\Vsd=\pm(\Delta+\Ea)/\mathrm{e}$ and the superconducting gap at $\Vsd=2\Delta/\mathrm{e}$.

\section{Sum Rule for Conductance Replicas}
\setcounter{myc}{7}
\begin{figure}[h]
	\includegraphics[width=0.5\columnwidth]{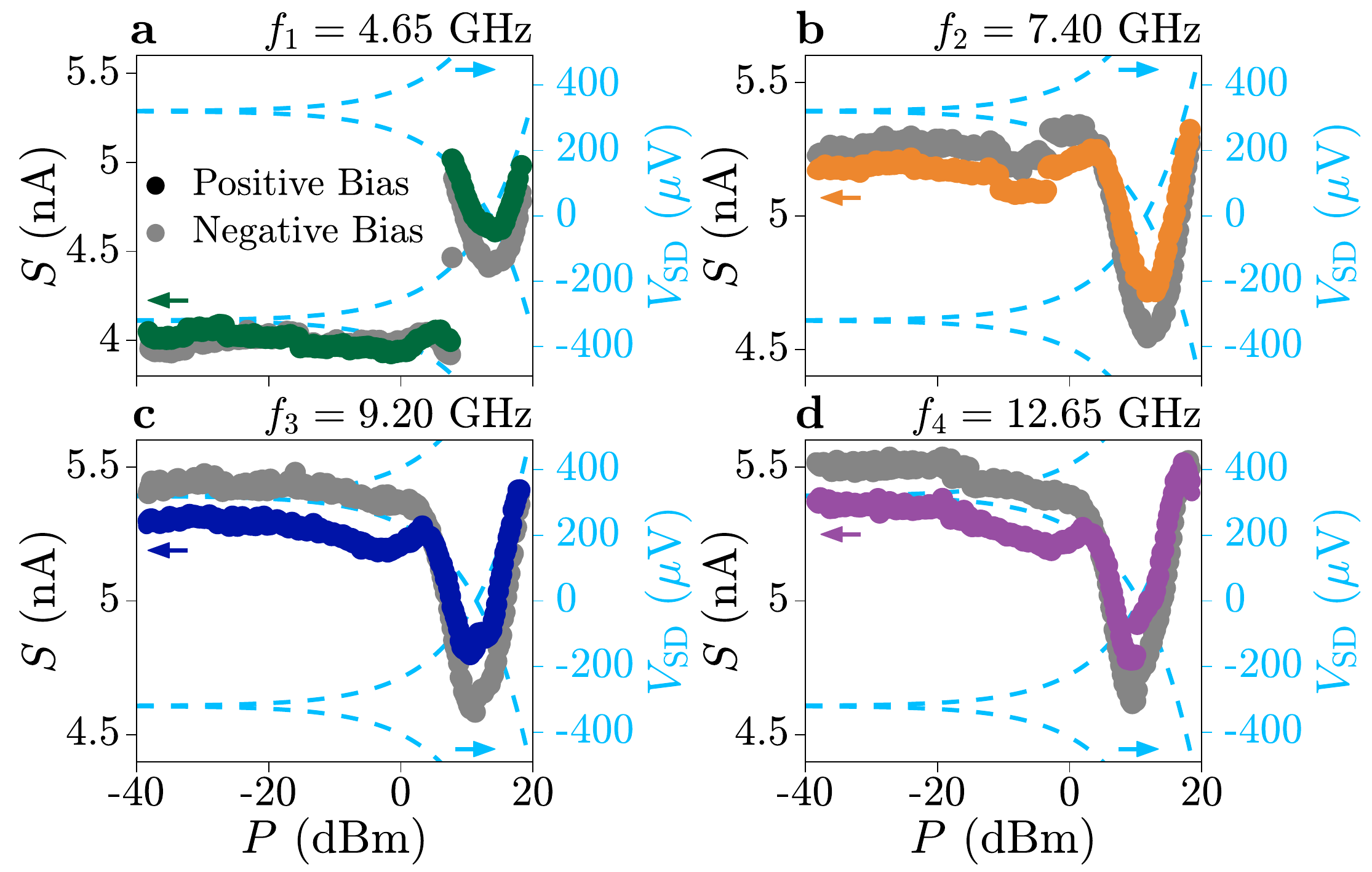}
	\caption{\textbf{Sum of conductance replicas under microwave irradiation of different frequencies.} \textbf{(a-d)} (Left axis) Sum of conductance over bias, $S$, as a function of power $P$ for the data in Figs.~2(g-j) of the Main Text, respectively. Frequencies $f_{i}$ correspond to those of Fig.~2 in the Main Text, where colours are defined. Coloured (grey) circles correspond to a sum over positive (negative) bias, $\Vsd>0$ ($\Vsd<0$). (Right axis) Dashed lines indicate the power dependence of conductance replicas, identical to Fig.~\ref{Sfig8}.}
	\label{Sfig15a}
\end{figure}

Reference~\cite{Park2022} described the importance of a sum-rule for conductance replicas to support their interpretation of Floquet-Andreev (F-A) states emerging under microwave irradiation. The sum rule brought forward in Ref.~\cite{Park2022} states that the sum of conductance over source-drain bias should be constant as a function of power, independent of the emergence of conductance replicas. This is expressed by the equation $S=\int_{0}^{\pm\infty}(\mathrm{d}I/\mathrm{d}V)\mathrm{d}V$, which is equivalent to a numerical integral of the experimental data. We applied the same technique to the results shown in Figs.~2(g-j) of the Main Text [see Fig.~\ref{Sfig15a}, data of Fig.~\ref{Sfig15a}(c) also plotted in inset of Fig.~3(c) of the Main Text]. The sum $S$ was calculated for each value of applied power $P$ by numerical integration of the differential conductance $G\equiv\mathrm{d}I/\mathrm{d}V$ over positive (negative) bias values, indicated by the coloured (grey) circles (left axis). Dashed lines in Fig.~\ref{Sfig15a} indicate the power dependence of high conductance features, as a function of bias $\Vsd$ (right axis). The power dependence is identical to those shown in Fig.~\ref{Sfig8}(a-d). The sum $S$ was approximately constant as a function of power up to $P\approx5~\mathrm{dBm}$. For $P\gtrapprox5~\mathrm{dBm}$, high conductance features were outside of the measurement range $-490~\mathrm{\mu V}<\Vsd<490~\mathrm{\mu V}$.  The change in $S$ was therefore consistent with conductance replicas exiting the measurement range, such that they were not included in $S$. The constant $S$ at low power is consistent with our conclusion that PAT was the dominant mechanism for conductance replicas, since it represents conservation of the number of states in the tunnel barrier and the junction. Equivalently, using Eq.~\ref{eqS2} we see that $S\propto\Sigma_{n}J^{2}_{n}(x)$ which is constant for a sum over all $n$. Hence, the total tunnel current through the barrier is constant as a function of power.

\section{Microwave Coupling Strength from High-Bias Conductance}
The coupling strength to the microwave field was calculated from the Shapiro steps in Fig.~\ref{Sfig15b}. We complement these values with calculations of the coupling strength directly from conductance features at high source-drain bias $\Vsd$. First, the background conductance [see Figs.~\ref{Sfig8}(a-d)] is fitted with a Gaussian function for each value of power $P$, or equivalently each value of microwave field amplitude $\Vmw$. Thus, values for the conductance peak position $V_{\mathrm{p}}(\Vmw)$ and standard deviation $\sigma(\Vmw)$ are obtained as a function of $\Vmw$. Then, the conductance peak position is fitted with a linear curve to obtain $V_{0}$. The values $V_{\mathrm{p}}(\Vmw)$ included in the fit are weighted by the standard deviations $\sigma(\Vmw)$. This method produces a value of $V_{0}$ for each microwave frequency $f$, along with an error $\delta V_{0}$ describing the uncertainty of the coupling strength to describe the data given the standard deviation $\sigma$. The obtained values are $V_{0}\pm\delta V_{0}=\{83.7\pm0.4,~97.6\pm1.5,~99.5\pm1.7,~100.6\pm2.9\}~\mathrm{\mu V}$, for frequencies $f=\{4.65,~7.20,~9.40,~12.65\}~\mathrm{GHz}$ respectively. The corresponding values of the dimensionless coupling strength are $\alpha_{0}=\{4.35\pm0.02,~3.19\pm0.05,~2.62\pm0.04,~1.92\pm0.06\}$. The standard deviation of the conductance peak was $60<\sigma<75~\mathrm{\mu V}$ for all datasets.

\section{Results for $V_{\mathrm{TG}}=-1.4~\mathrm{V}$}
Conductance replication demonstrated in the Main Text was obtained at $\Vtg=-0.8~\mathrm{V}$. Here we show measurements on the same device at $\Vtg=-1.4~\mathrm{V}$. Figures~\ref{Sfig2}-\ref{Sfig5} show bias-spectroscopy as a function of applied microwave power, for tunnel gate voltages $\Vt=-2.06~\mathrm{V},~-2.08~\mathrm{V},~-2.1~\mathrm{V}~\mathrm{and}~-2.12~\mathrm{V}$ respectively. For high tunnel barrier transparency [$\Vtg=-2.06~\mathrm{V}$, Fig.~\ref{Sfig2}], a conductance peak at $\Vsd=0$ was indicative of a supercurrent flowing across the tunnel barrier. On increasing applied microwave power, conductance replicas emerged in both the low and high bias features, at the same power and with the same dependence. For lower tunnel barrier transparency [Figs.~\ref{Sfig3}-\ref{Sfig5}], conductance features at high bias were replicated with separation $\Delta\Vsd=hf/\mathrm{e}$. The mean separation of low and high bias replicas are displayed as filled grey squares and circles in Fig.~3(b) of the Main Text, respectively. The bias separation of conductance replicas was consistent with $\Delta\Vsd=hf/q$, where $q$ is the charge tunnelling across the barrier. As in the Main Text, concurrent replicas in low and high bias features indicated PAT as the dominant mechanism. 

The coupling strength to the microwave field is calculated for this $\Vtg$ value from conductance features at high source-drain bias in Fig.~\ref{Sfig4}, using the same procedure as outlined in the previous section. The obtained values are $V_{0}\pm\delta V_{0}=\{80.6\pm2.3,~100.2\pm2.6,~99.1\pm2.7,~96.0\pm2.3\}~\mathrm{\mu V}$, for frequencies $f=\{4.65,~7.20,~9.40,~12.65\}~\mathrm{GHz}$ respectively. The corresponding values of the dimensionless coupling strength are $\alpha_{0}=\{4.19\pm0.12,~3.27\pm0.08,~2.60\pm0.07,~1.84\pm0.04\}$. As for the data taken at $\Vtg=-0.8~\mathrm{V}$, the standard deviation of the conductance peak was $60<\sigma<75~\mathrm{\mu V}$ for all datasets.

The values for $\alpha_{0}$ at $\Vtg=-1.4~\mathrm{V}$ show remarkable agreement with those at $0.8~\mathrm{V}$ [see blue lines in Figs.~\ref{Sfig4}(c-f)]. The change in coupling strength as a result of the more negative $\Vtg$ is quantified by $\Delta\alpha_{0}\equiv\alpha_{0}(\Vtg=-0.8)-\alpha_{0}(\Vtg=-1.4)=\{-0.16\pm0.12,~0.08\pm0.09,~-0.02\pm0.08,~0.08\pm0.07\}$ for the respective frequencies of applied microwave radiation. Uncertainties are calculated from the sum over variances of each $\alpha_{0}$ value. These results show that the change in coupling strength as a result of the more negative gate voltage was at most $4\%$.

\begin{figure}[h]
	\includegraphics[width=0.5\columnwidth]{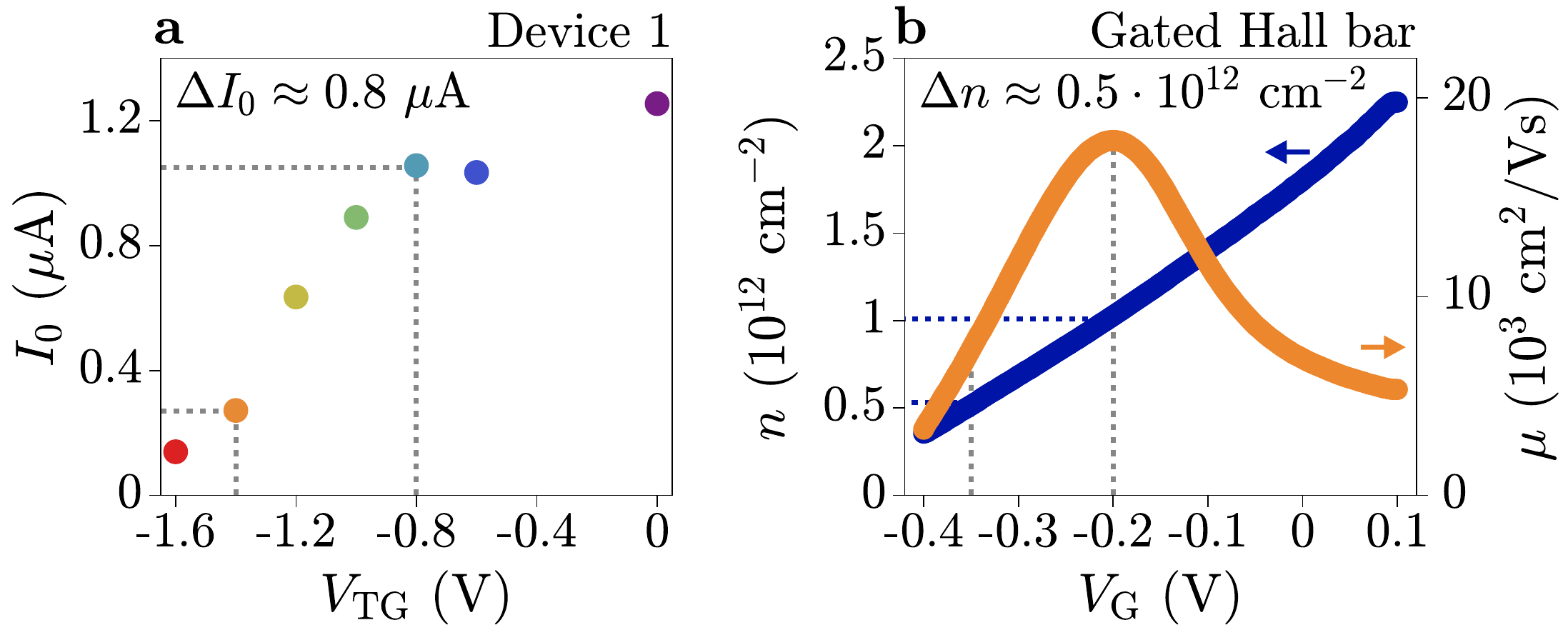}
	\centering
	\caption{\textbf{Estimating the change in carrier density as a function of $\Vtg$.} (a) Maximum switching current $I_{0}$ in Device 1 as a function of top-gate voltage $\Vtg$. Data points at $\Vtg=-0.8~\mathrm{V}$ and $-1.4~\mathrm{V}$ are indicated by dashed lines, corresponding to a difference $\Delta I_{0}\approx0.8~\mathrm{\mu A}$. (b) Measurements of a gated Hall bar in the same material. Carrier density $n$ (blue, left axis) and mobility $\mu$ (orange, right axis) are plotted as a function of the global gate voltage $V_{\mathrm{G}}$. Dashed lines indicate the estimated change in carrier density as $\Delta n\approx 0.5\cdot10^{12}~\mathrm{cm^{-2}}$.}
	\label{Sfig19}
\end{figure}

\begin{figure}[h]
	\includegraphics[width=0.5\columnwidth]{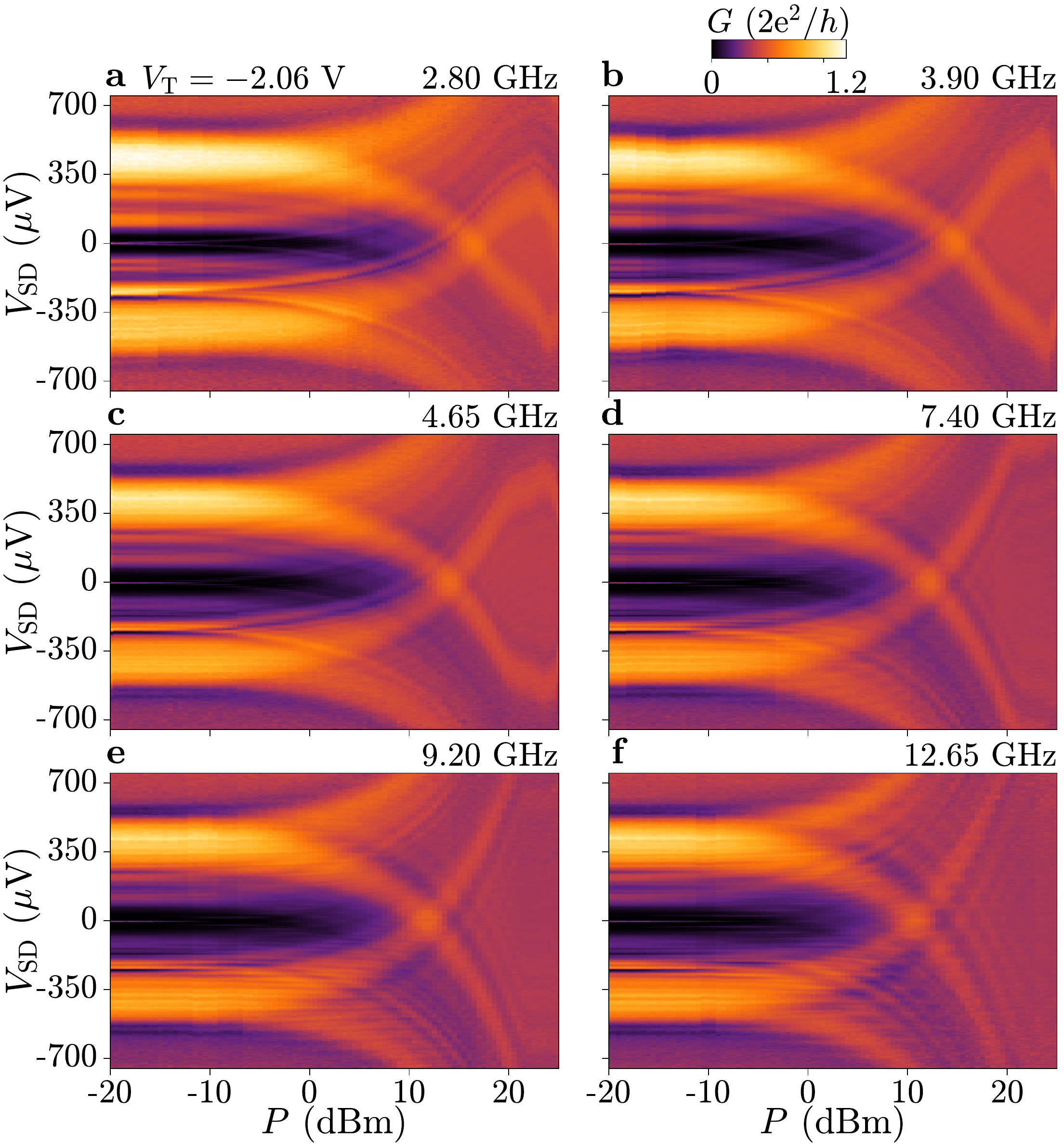}
	\caption{\textbf{Power dependence at $\Vtg=-1.4~\mathrm{V}$ and $\Vt=-2.06~\mathrm{V}$.} \textbf{(a-f)} Conductance $G$ as a function of source-drain bias $\Vsd$ and power $P$, for frequencies ${f=\left\{2.80,~3.90,~4.65,~7.40,~9.20,~12.65\right\}~\mathrm{GHz}}$. Mean separation of replicated supercurrent features is shown in Fig.~3(b) of the Main Text (full grey squares).}
	\label{Sfig2}
\end{figure}

\begin{figure}
	\includegraphics[width=0.5\columnwidth]{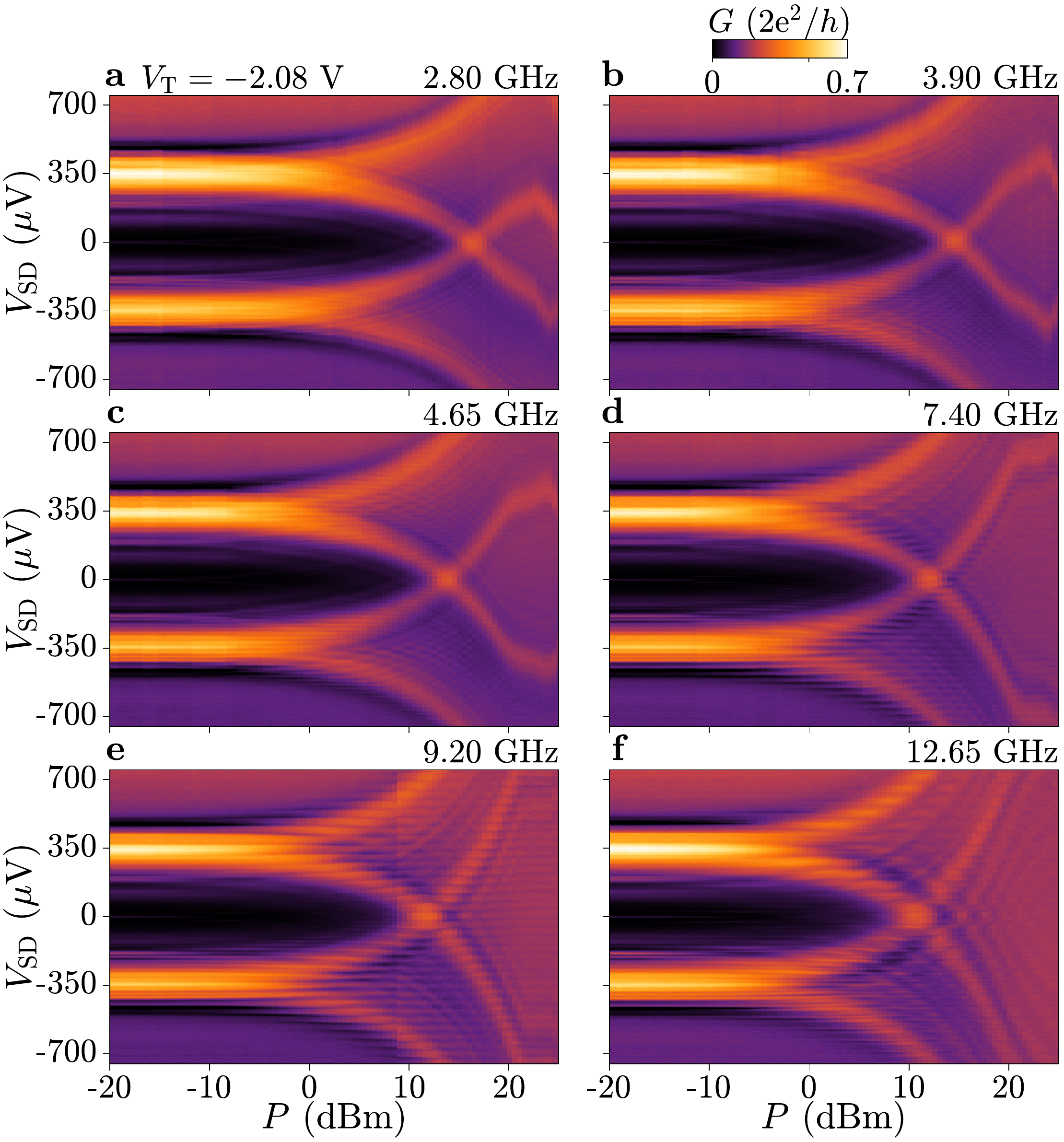}
	\caption{Same as \ref{Sfig2} for $\Vt=-2.08~\mathrm{V}$.}
	\label{Sfig3}
\end{figure}

\begin{figure}
	\includegraphics[width=0.5\columnwidth]{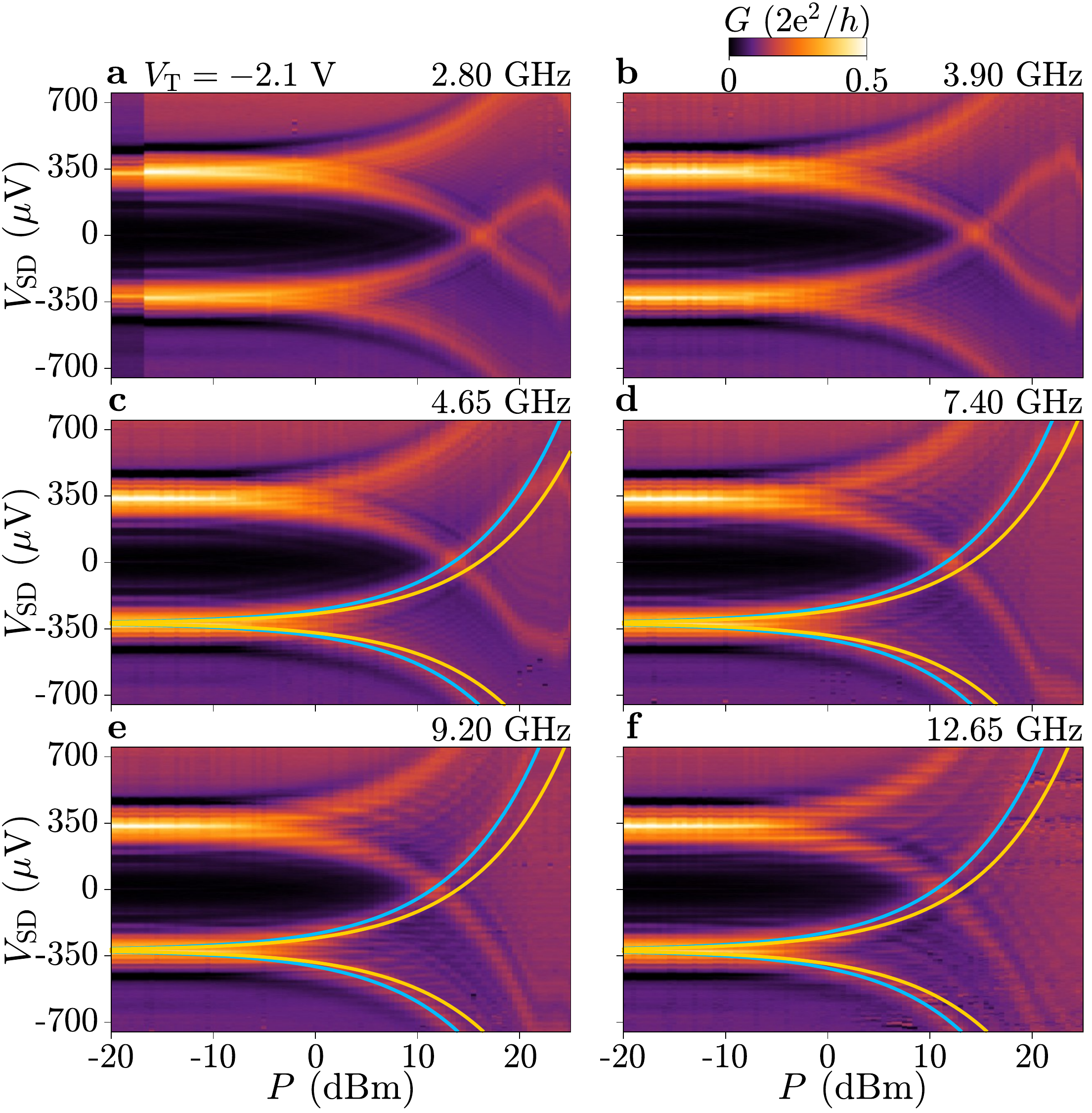}
	\caption{Same as \ref{Sfig2} for $\Vt=-2.1~\mathrm{V}$. Mean separation of replicated conductance features is shown in the Fig.~3(b) of the Main Text (full grey circles). Power dependence of conductance replicas obtained for $\Vtg=-0.8~\mathrm{V}$ [blue lines, identical to Fig.~\ref{Sfig8}] is plotted in (c-f), alongside the expectation for a $25\%$ decrease in coupling strength due to smaller Fermi velocity [yellow lines].}
	\label{Sfig4}
\end{figure}

\begin{figure}
	\includegraphics[width=0.5\columnwidth]{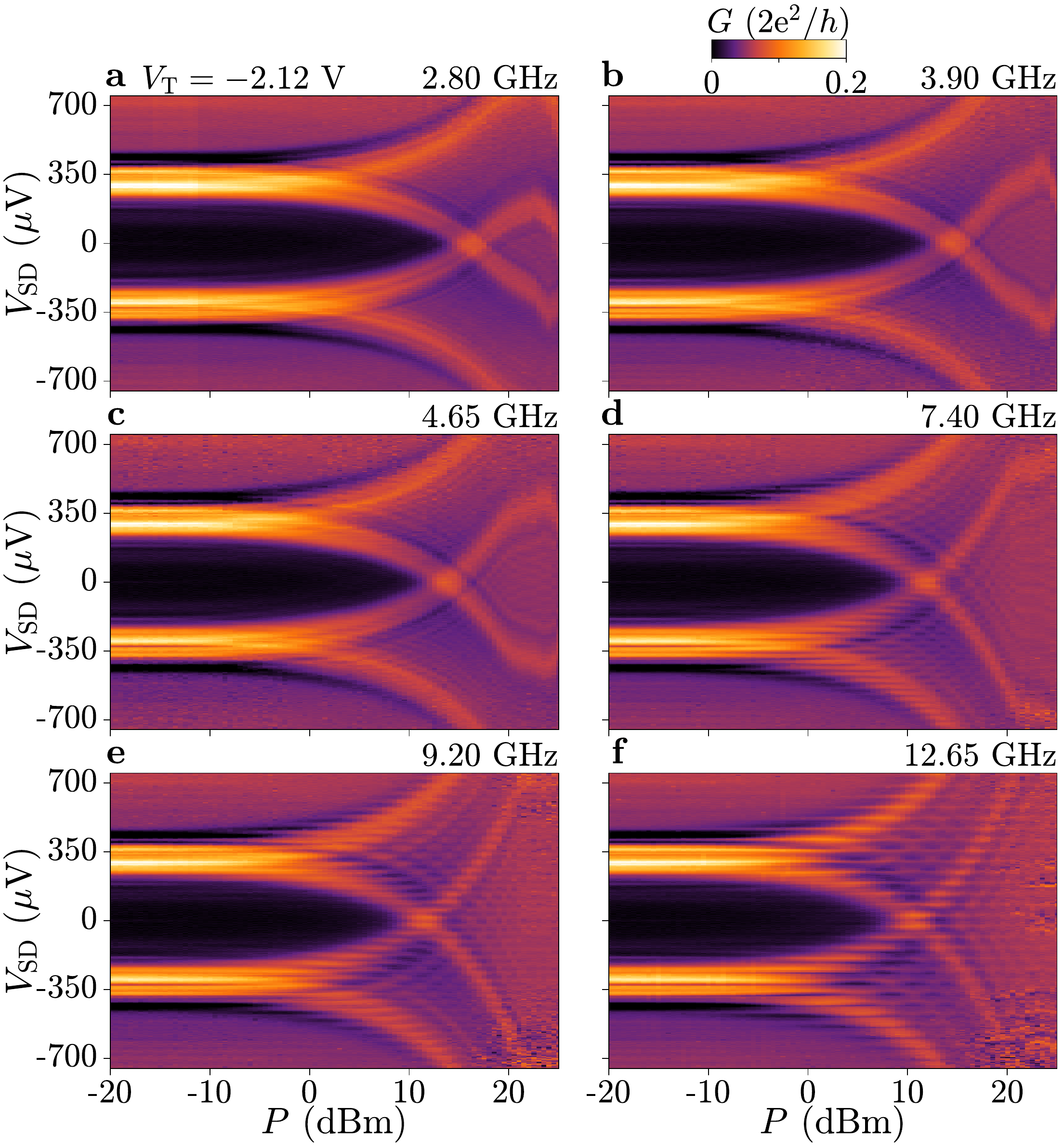}
	\caption{Same as \ref{Sfig2} for $\Vt=-2.12~\mathrm{V}$.}
	\label{Sfig5}
\end{figure}

The carrier density in the SNS junction is expected to change as a function of $\Vtg$. The maximum switching current $I_{0}$ of Device~1 is plotted in Fig.~\ref{Sfig19}(a) as a function of $\Vtg$. Data points corresponding to $\Vtg=-0.8~\mathrm{V}$ and $-1.4~\mathrm{V}$ are indicated by dashed lines, and show that $I_{0}$ at $\Vtg=-1.4~\mathrm{V}$ was $\sim25\%$ the value at $\Vtg=-0.8~\mathrm{V}$. The change in maximum switching current was $\Delta I_{0}\approx0.8~\mathrm{\mu A}$. At $\Vtg=-0.8~\mathrm{V}$, $I_{0}$ reached a peak after a linear increase from the most negative $\Vtg$ values. We associate this linear regime to occupation of the first subband in the semiconductor. The open ($\Vtg=-0.8~\mathrm{V}$) and closed ($\Vtg=-1.4~\mathrm{V}$) regimes are therefore associated with full and partial occupation of the first subband, respectively. Gated Hall bar measurements in the same material are shown in Fig.~\ref{Sfig19}(b). The density $n$ and mobility $\mu$ are plotted as a function of the gate voltage $V_{\mathrm{G}}$. The gate lever arm was different in the Hall bar and SNS junction due to different fabrication processes for each chip. Therefore, we estimate the change in density from the range of single subband occupation, where the mobility $\mu$ increased linearly with $n$. The carrier density at peak $\mu$ was compared to that where the mobility was $25\%$ above its lowest measured value. This was chosen to approximately correspond to the $I_{0}$ value at $\Vtg=-1.4~\mathrm{V}$ relative to $\Vtg=-0.8~\mathrm{V}$. This gave an approximate change in carrier density of $\Delta n\approx0.5\cdot10^{12}~\mathrm{cm^{-2}}$, or $\Delta n/n\approx0.5$. We therefore estimate a $25\%$ decrease in the Fermi velocity for $\Vtg=-1.4~\mathrm{V}$ relative to $\Vtg=-0.8~\mathrm{V}$. While this value is an approximation, the large change in $I_{0}$ is indicative of an appreciable change in the carrier density.

From the theory of Floquet-Andreev states~\cite{Park2022}, a $25\%$ decrease in the Fermi velocity would correspond to a $25\%$ decrease in the microwave coupling strength $\alpha_{0}$. This is plotted as the yellow lines in Figs.~\ref{Sfig4}(c-f), and does not match the experimental result. To be consistent with a Floquet-Andreev interpretation, calculated values of $\Delta\alpha_{0}$ imply a change in the Fermi velocity of less that $4\%$, incompatible with switching current and Hall bar measurements, or an alternative mechanism which almost exactly compensates for the change in carrier density. In contrast, no gate dependence is expected in the PAT interpretation. This further supports PAT as the dominant mechanism for conductance replicas.

\setcounter{myc}{8}

\setcounter{myc}{9}

\setcounter{myc}{10}

\setcounter{myc}{11}

\setcounter{myc}{12}

\pagebreak

\section{Conductance Replication in a Second Device}
\setcounter{myc}{13}
\begin{figure}[h]
	\includegraphics[width=0.5\columnwidth]{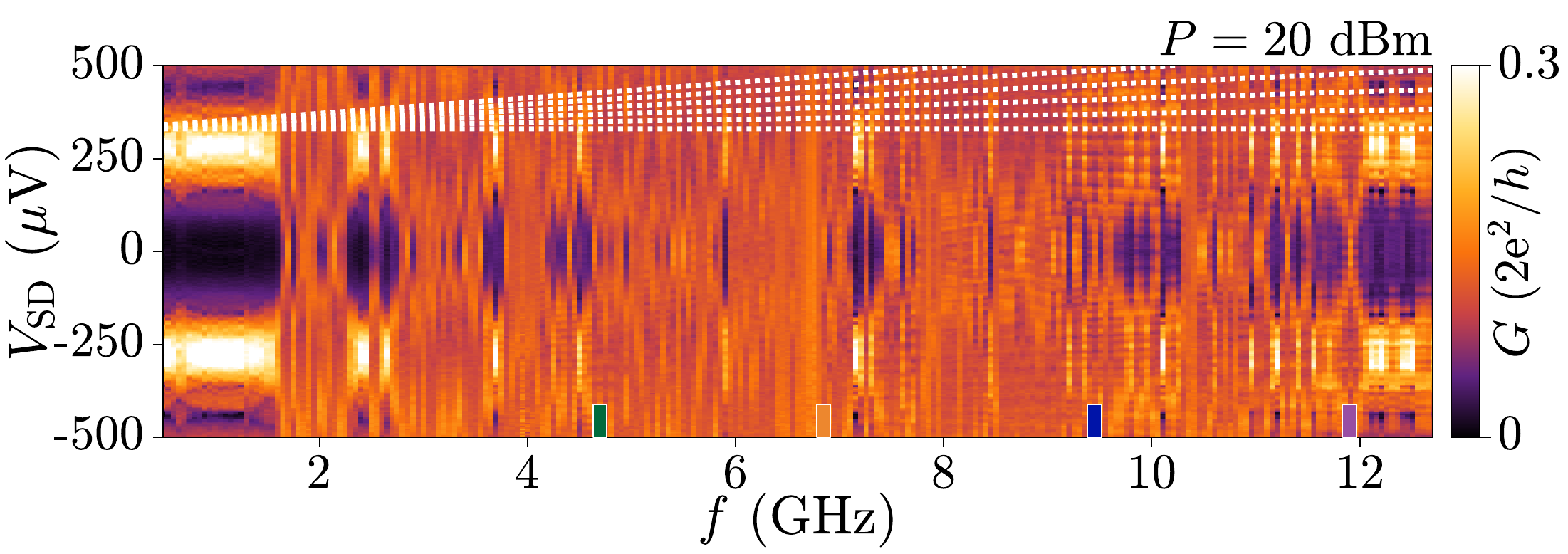}
	\caption{\textbf{Frequency dependence of conductance $G$ as a function of source-drain bias $\Vsd$ in Device 2, at fixed power $P=20~\mathrm{dBm}$.} Conductance replicas are schematically indicated by the dashed white line, $\Delta\Vsd=hf/\mathrm{e}$. Coloured markers indicate the frequencies used in Figs.~\ref{Sfig13} and \ref{Sfig14}.}
	\label{Sfig11}
\end{figure}

\setcounter{myc}{14}
\begin{figure}
	\includegraphics[width=0.5\columnwidth]{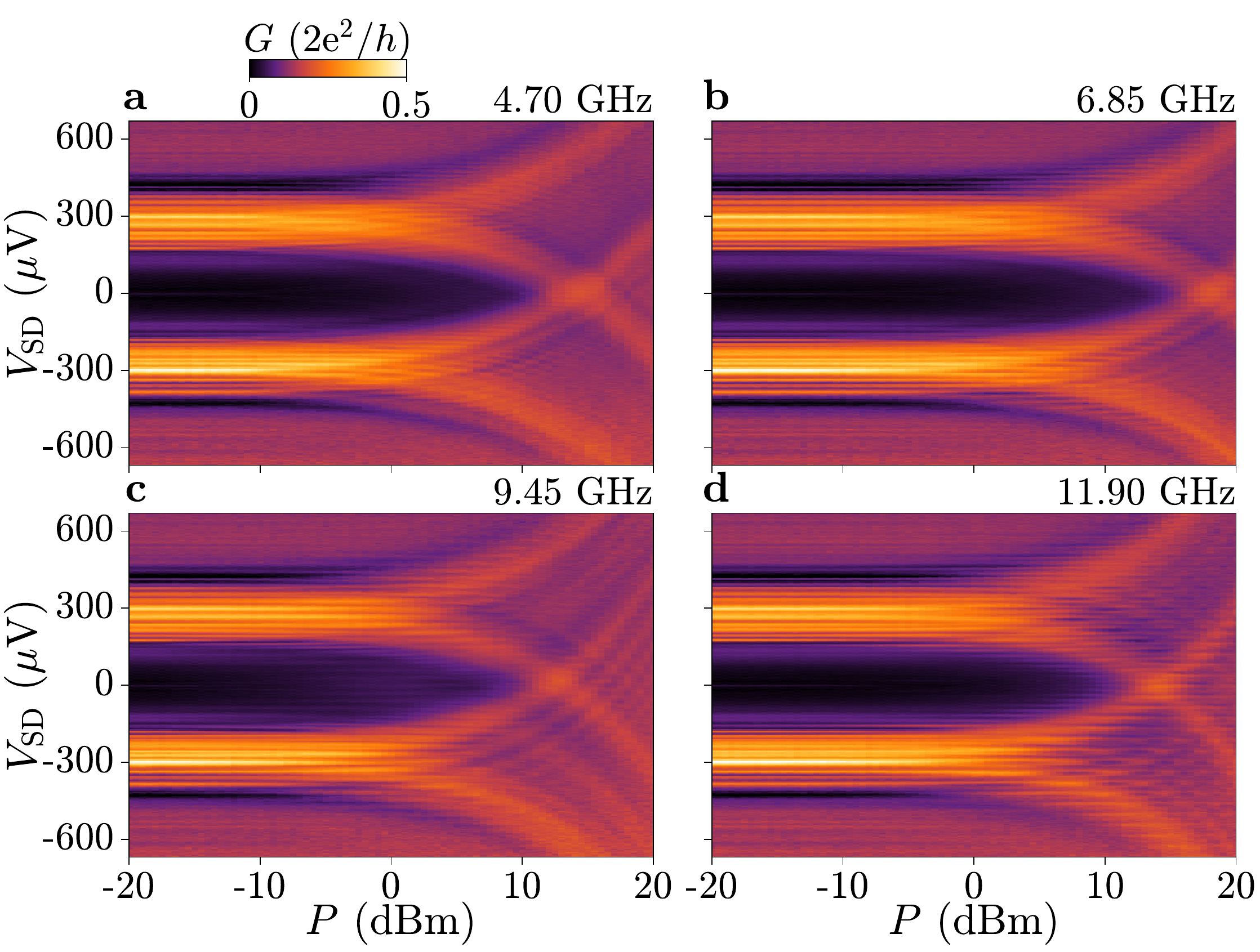}
	\caption{\textbf{Conductance of Device 2 as a function of applied microwave power $P$, for frequencies ${f=\{4.70,~6.85,~9.45,~11.90\}~\mathrm{GHz}}$ in (a-d) respectively.}}
	\label{Sfig13}
\end{figure}

\setcounter{myc}{15}
\begin{figure*}
	\includegraphics[width=\textwidth]{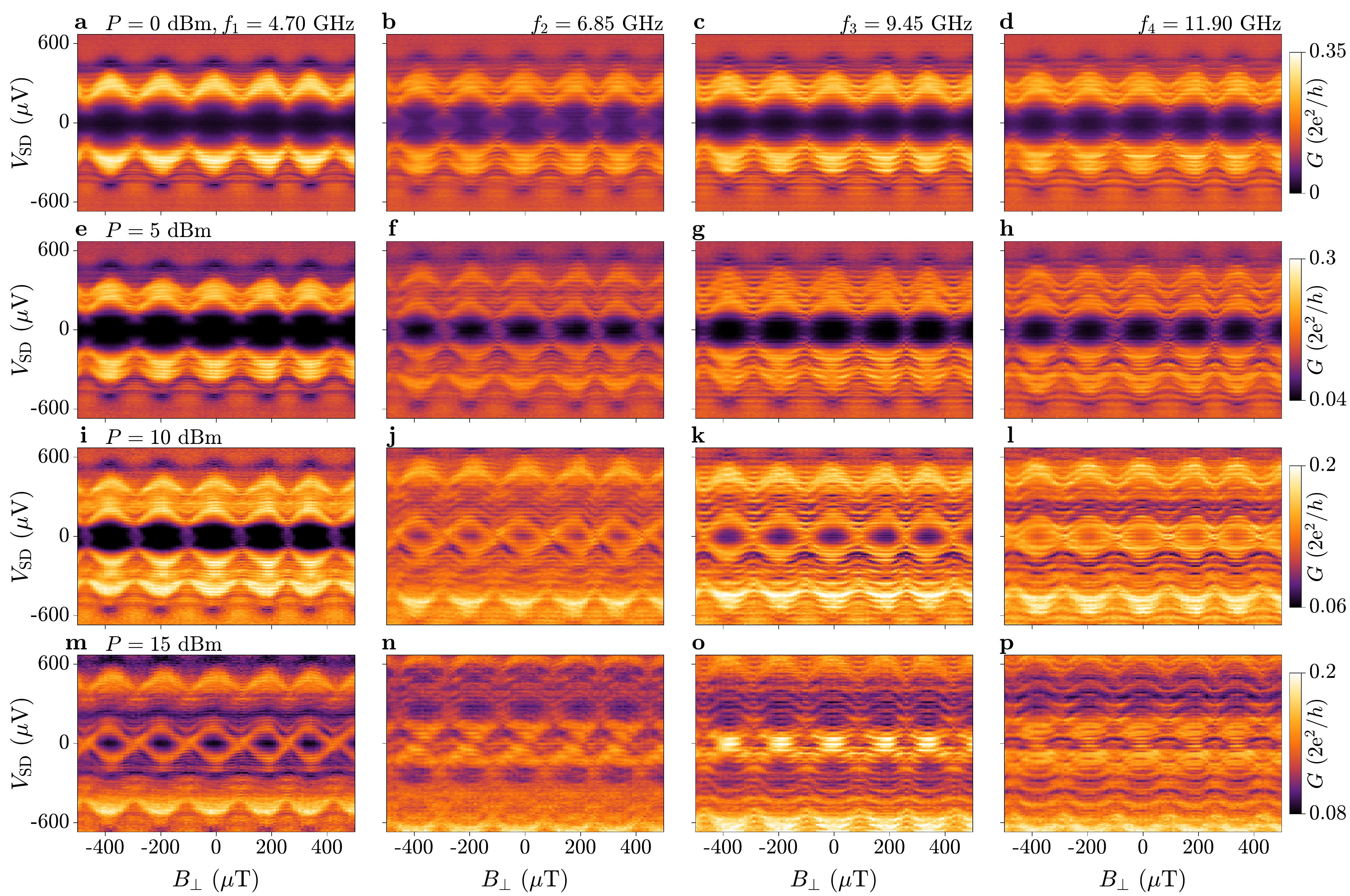}
	\caption{\textbf{Conductance of Device 2 as a function of perpendicular magnetic field $\Bperp$ for different frequencies $f$ and powers $P$.} \textbf{(a-d)} $\Bperp$ dependence of conductance response at $P=0~\mathrm{dBm}$ for frequencies $f=\{4.70,~6.85,~9.45,~11.90\}~\mathrm{GHz}$, respectively. \textbf{(e-h)} Same as (a-d) for $P=5~\mathrm{dBm}$. \textbf{(i-l)} Same as (a-d) for $P=10~\mathrm{dBm}$. \textbf{(m-p)} Same as (a-d) for $P=15~\mathrm{dBm}$.}
	\label{Sfig14}
\end{figure*}
Measurements were performed on a second device, fabricated on the same chip and lithographically similar to the first except for the width of the SNS junction, which was $500~\mathrm{nm}$ rather than $2.5~\mathrm{\mu m}$ for Device 1. Measurements are shown for tunnel gate voltages $\Vt=-0.768~\mathrm{V}$ a top gate voltage $\Vtg=0~\mathrm{V}$, kept constant throughout the measurements. Figure~\ref{Sfig11} shows the frequency response of Device 2 to microwave irradiation at an applied power of $P=20~\mathrm{dBm}$. The frequency response was similar to that of Device 1 [see Fig.~\ref{Sfig1}], showing conductance replicas with separation $\Delta\Vsd=hf/\mathrm{e}$ indicated by the white dashed lines. Frequencies $f=4.70~\mathrm{GHz},~6.85~\mathrm{GHz},~9.45~\mathrm{GHz}~\mathrm{and}~11.90~\mathrm{GHz}$ are indicated by the coloured markers, where many replicas are evident. Figure~\ref{Sfig13} shows the conductance response to microwave irradiation at these frequencies, for increasing microwave power $P$. Conductance replicas emerged with separation $\Delta\Vsd=hf/\mathrm{e}$, shown as empty grey circles in Fig.~3(b) of the Main Text. Figure~\ref{Sfig14} shows the conductance as a function of perpendicular magnetic field $\Bperp$, for increasing microwave power. Field-periodic conductance features were replicated, with more replicas emerging for increasing applied power. 

\begin{figure*}
	\includegraphics[width=0.5\columnwidth]{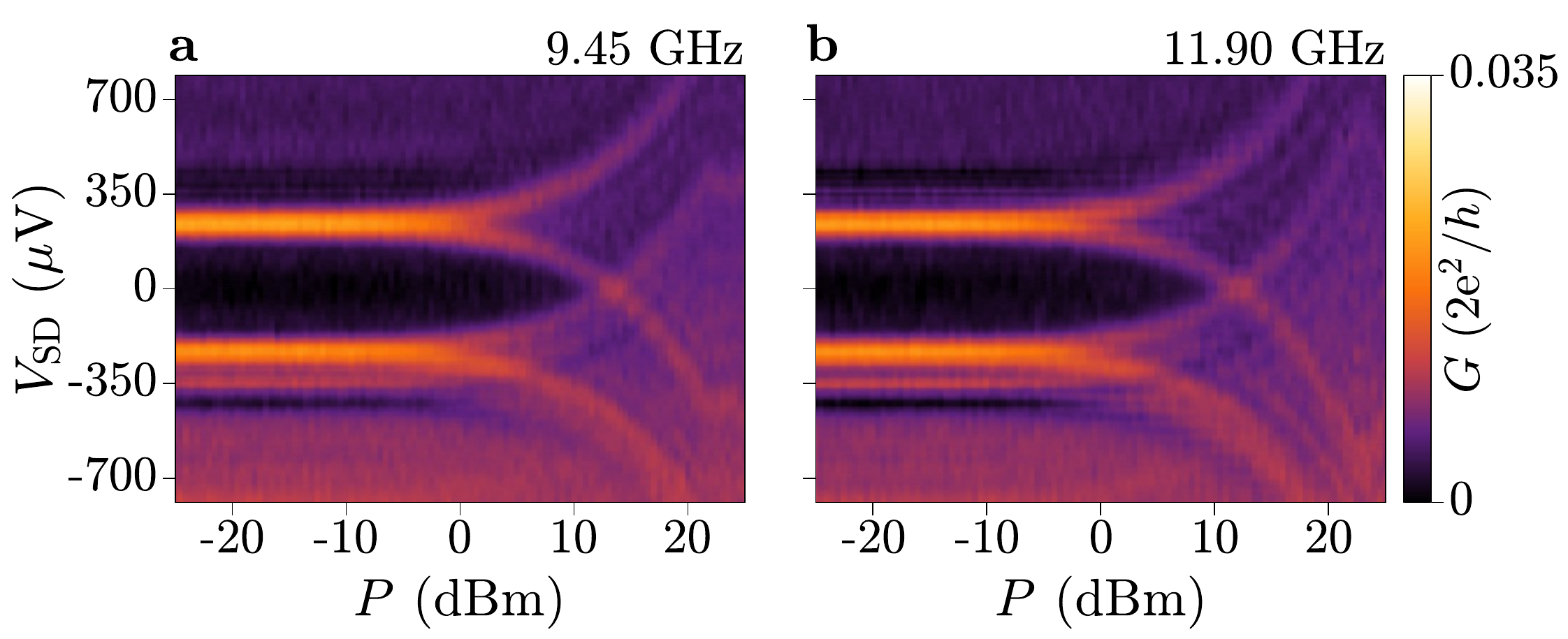}
	\caption{\textbf{Conductance replicas in Device 2 at low tunnel-barrier transparency $(V_{\mathrm{T,L}},~V_{\mathrm{T,R}})=(-0.911,~-0.875)~\mathrm{V}$, for frequencies $9.45~\mathrm{GHz}$ and $11.90~\mathrm{GHz}$ in (a,~b) respectively.}}
	\label{Sfig18}
\end{figure*}

Figure~\ref{Sfig18} shows the differential conductance as a function of applied power when the transparency of the tunnel barrier was significantly reduced, by setting tunnel gate voltages to $(V_{\mathrm{T,L}},~V_{\mathrm{T,R}})=(-0.911,~-0.875)~\mathrm{V}$. Conductance replicas emerge up to large applied powers, as in Fig.~\ref{Sfig13}. 

\setcounter{myc}{16}

\section{$B_{\perp}$-Dependence in Spectroscopy}
\setcounter{myc}{17}
\begin{figure}[h]
	\includegraphics[width=0.5\columnwidth]{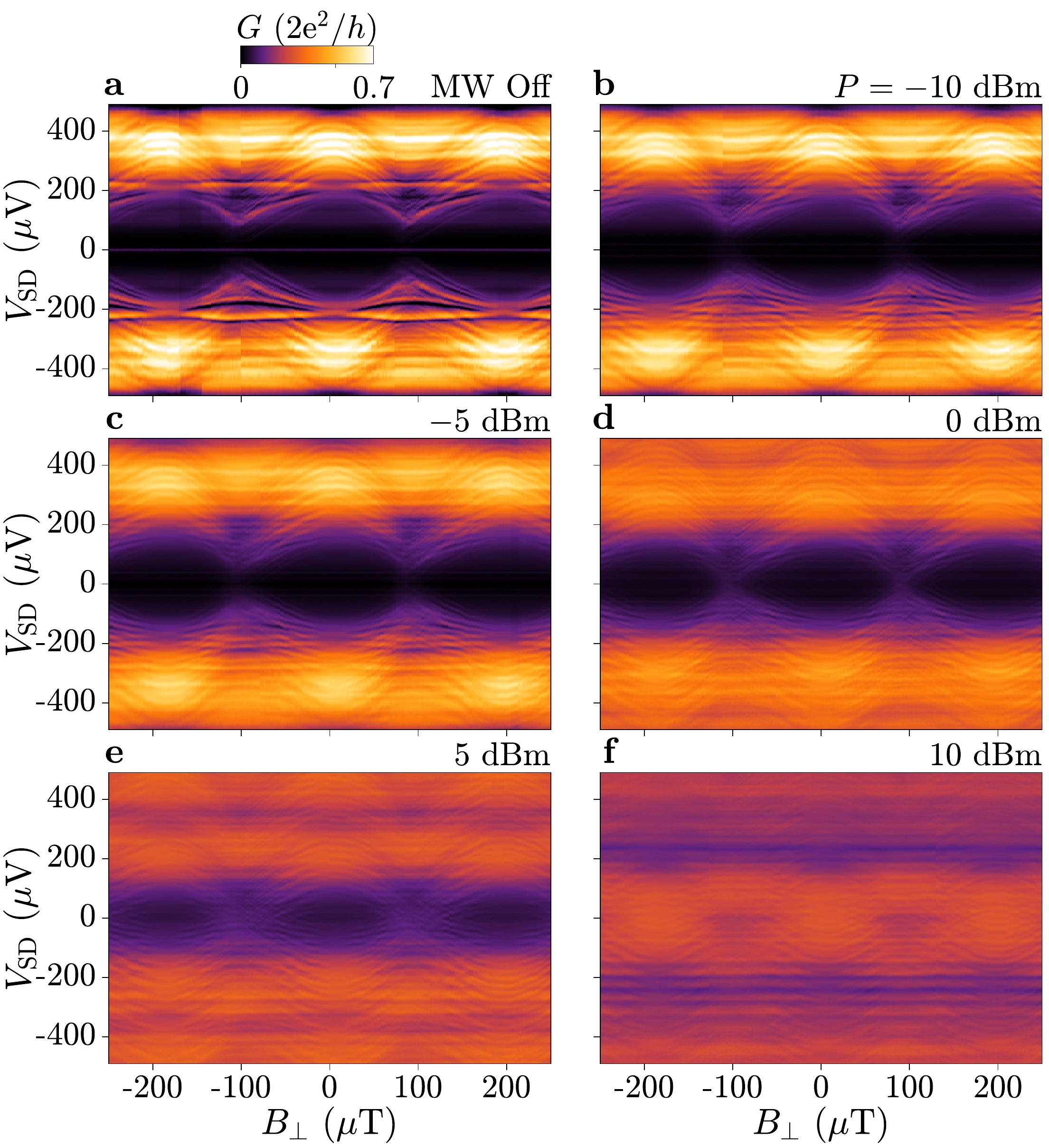}
	\caption{\textbf{Conductance as a function of perpendicular magnetic field $\Bperp$.} \textbf{(a)} Conductance $G$ as a function of $\Bperp$ with no microwave field applied. \textbf{(b-f)} Conductance $G$ as a function of $\Bperp$ under microwave irradiation at frequency $f=9.20~\mathrm{GHz}$. Applied microwave powers of ${P=\{-10,~-5,~0,~5,~10\}~\mathrm{dBm}}d\varphi$, respectively. Same gate configuration as Fig. 4 in the Main Text.}
	\label{Sfig6}
\end{figure}

Selected conductance maps as a function of perpendicular magnetic field $\Bperp$ are shown in Figs.~4(a-c) of the Main Text. The full dataset is shown in Fig.~\ref{Sfig6}, for no applied microwaves [Fig.~\ref{Sfig6}(a)] and applied powers ranging from $P=-10~\mathrm{dBm}$ [Fig.~\ref{Sfig6}(b)] to $P=10~\mathrm{dBm}$ [Fig.~\ref{Sfig6}(f)]. Some conductance features were periodic in $\Bperp$; these corresponded to ABSs in the SNS junction, which were dependent on the phase difference across the junction. Field-independent features corresponded to the superconducting gap edge at $\Vsd=2\Delta/\mathrm{e}$, and conductance resonances in the tunnelling probe. For increasing microwave power, additional field-periodic features appeared in the conductance map. The magnitude of the conductance at a given bias decreased, as it was distributed across more conductance peaks. This is consistent with current conservation in the PAT process.

\section{Switching Current of the Planar SQUID}
\setcounter{myc}{18}
\begin{figure}[h]
	\includegraphics[width=0.5\columnwidth]{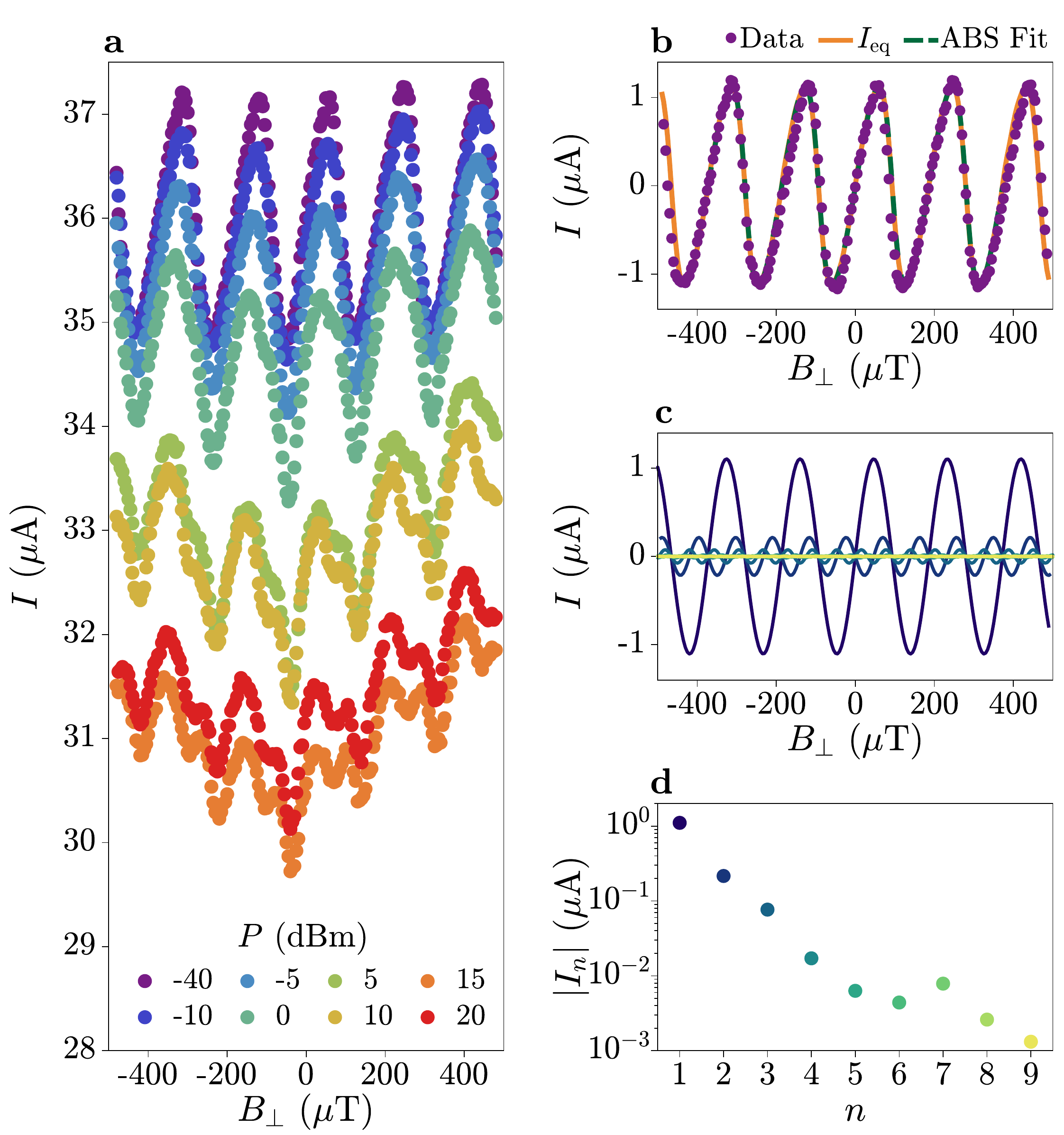}
	\caption{\textbf{Current phase relation under microwave irradiation at $f=9.20~\mathrm{GHz}$.} \textbf{(a)} Switching current $I$ of the planar SQUID as a function of $\Bperp$, as a function of power $P$. \textbf{(b)} CPR data at $P=-40~\mathrm{dBm}$, after subtraction of background current corresponding to Al constriction (circles). Equilibrium current $I_{\mathrm{eq}}$ determined from harmonics $I_{n}$ up to tenth order (orange line). Fit to CPR using Eq.~\ref{eq2} (green dashed line), giving an effective transmission of $\bar{\tau}=0.84$. \textbf{(c)} Harmonics $I_{n}\sin(n\varphi)$ extracted from the low power CPR up to the tenth order. Colours are defined in (d). \textbf{(d)} Absolute amplitude $|I_{n}|$ of the harmonics plotted in (c).}
	\label{Sfig12}
\end{figure}

Current-biased measurements were performed by applying a current $\Idc$ to a low-impedance superconducting lead on the right side of the planar SQUID loop, which flows to ground via a second low-impedance superconducting lead at the bottom of the device [see Fig.~1(c) of Main Text]. The current was prevented from flowing through the probe by floating its contacts. The differential voltage drop across the planar SQUID, $\Vtwo$, was measured to detect the switching current to the resistive state. The SNS junction was embedded in a superconducting loop defined by a $400~\mathrm{nm}$ wide epitaxial Al stripe enclosing an area of $A=10~\mathrm{\mu m^{2}}$. The width of a $200~\mathrm{nm}$ portion of the loop was reduced to $130~\mathrm{nm}$, which reduced the switching current of the loop from several $\mathrm{mA}$ to $36~\mathrm{\mu A}$. Without this Al constriction, the switching current background would be too large to be measured without dissipating large amounts of heat at the mixing chamber of the fridge. The switching current of the Al constriction was still more than a factor of 30 larger than the switching current of the SNS junction. Due to the large asymmetry in the critical currents of the planar SQUID, the oscillations correspond to the current-phase relation (CPR) of the SNS junction and the background to the Al constriction. Hence, a perpendicular magnetic field $\Bperp$ applied to the loop of area $A$ resulted in a phase drop of $\varphi=2\mathrm{\pi}\Bperp A/\mathit{\Phi}_{0}$ across the SNS junction. 

The switching current of the planar SQUID is shown in Fig.~\ref{Sfig12}(a), for increasing microwave power $P$. At low power [purple circles], oscillations with a period of $B_{\mathrm{Period}}\approx200~\mathrm{\mu T}$ and peak-to-peak amplitude of $2~\mathrm{\mu A}$ were observed, on top of a constant background of $36~\mathrm{\mu A}$. For increasing applied power, the amplitude of the oscillations decreased and their shape was distorted (as described in the Main Text), while the background switching current decreased and developed a pronounced minimum close to $\Bperp=0$. The decrease in the switching current of the constriction under microwave irradiation is assigned to pair-breaking in the Al by photon absorption, which may also account for the enhanced switching current suppression close to $\Bperp=0$ by quasiparticle generation in the constriction and the superconducting leads~\cite{Peltonen2011}.

The CPR of the SNS junction was obtained by subtracting the switching current of the constriction, as shown in Fig.~\ref{Sfig12}(b) for $P=-40~\mathrm{dBm}$ [purple circles]. The switching current of the constriction was found by a polynomial fit to the data across six full periods, such that the resulting CPR was symmetric with respect to current and had a constant oscillation amplitude over all periods. The microwave field did not affect the switching current at this low power, so the CPR is considered to be at equilibrium. We described the data by extracting the harmonics up to the $10^{\mathrm{th}}$ order, using the following equation:

\setcounter{myc2}{4}
\begin{equation}
	I_{\mathrm{eq}} = \Sigma_{n}I_{n}\sin(n\varphi),
	\label{eq1}
\end{equation}
where $I_{n}=(1/\mathrm{\pi})\int_{0}^{2\mathrm{\pi}}I_{eq}\sin(n\varphi)\mathrm{d}\varphi$.

The equilibrium supercurrent is plotted as the orange line in Fig.~\ref{Sfig12}(b), composed of the harmonics in Fig.~\ref{Sfig12}(c) with amplitudes $|I_{n}|$ plotted in Fig.~\ref{Sfig12}(d). The presence of $n>1$ terms, which gives the forward skewness of the CPR, is indicative of the presence of highly transmissive ABSs in the junction~\cite{Beenakker1991,Spanton2017,Nichele2020}. Since these ABSs carry the supercurrent, the CPR is described in terms of the ABS properties. However, the junction contained many modes, each with a distinct transmission $\tau$, which all contribute to the supercurrent. It was not feasible to assign a transparency to each individual mode, so we instead considered a junction where all modes have an equal effective transmission $\bar{\tau}$. This describes the macroscopic properties of the junction, but does not capture details of the individual microscopic states. The CPR was then described by

\setcounter{myc2}{5}
\begin{equation}
	I_{\mathrm{ABS}} = I_{0}\frac{\bar{\tau}\sin(\varphi)}{\Ea(\varphi)/\Delta},
	\label{eq2}
\end{equation}
where $\Ea=\Delta\sqrt{1-\bar{\tau}\sin^{2}(\varphi/2)}$ is the ABS energy and $I_{0} = (\mathrm{e}/2\hbar)\bar{N}\Delta$, where $\bar{N}$ is the effective number of modes in the junction. A fit to the low power data gave $\bar{\tau}=0.84$ [green dashed line in Fig.~\ref{Sfig12}(b)], consistent with the presence of highly transmissive modes observed in tunnelling spectroscopy (see Fig.~1(e) in the Main Text).

\section{Adiabatic Theory of the Current-Phase Relation under Microwave Irradiation}
We use an adiabatic theory of an SNS junction under microwave irradiation to describe the CPR under increasing applied power~\cite{Barone1982,Bergeret2011,Dou2021}. A monochromatic drive at frequency $f$ generates a time-varying voltage ${V(t)=\Vmw\sin(2\mathrm{\pi} ft)}$, resulting in a time-varying phase across the SNS junction of $\varphi(t) = \varphi_{0}+2\alpha\cos(2\mathrm{\pi} ft)$. The electromagnetic field strength is described by the parameter $\alpha=\mathrm{e}\Vmw/hf$. In the adiabatic approximation, the stationary phase at equilibrium [Eq.~\ref{eq1}] is replaced by the time-varying phase $\varphi(t)$. No excitation of ABSs is considered in this model. The resulting CPR is:

\setcounter{myc2}{6}
\begin{equation}
	I_{\mathrm{ad.}} = \Sigma_{n}I_{n}J_{0}(2n\alpha)\sin(n\varphi),
\end{equation}
where $J_{0}$ is a zero-order Bessel function of the first kind and $I_{n}$ are the harmonic coefficients obtained for the equilibrium CPR. The CPR traces under microwave irradiation were therefore fitted with $\alpha$ as a single free parameter, using the $I_{n}$ shown in Fig.~\ref{Sfig12}(d). The results of the fit are shown in Fig.~4(e) of the Main Text.

%\section{Fitting of Microwave Field Strength}

\section{Non-thermal ABS Occupation}
\setcounter{myc}{19}
\begin{figure}[h]
	\includegraphics[width=0.5\columnwidth]{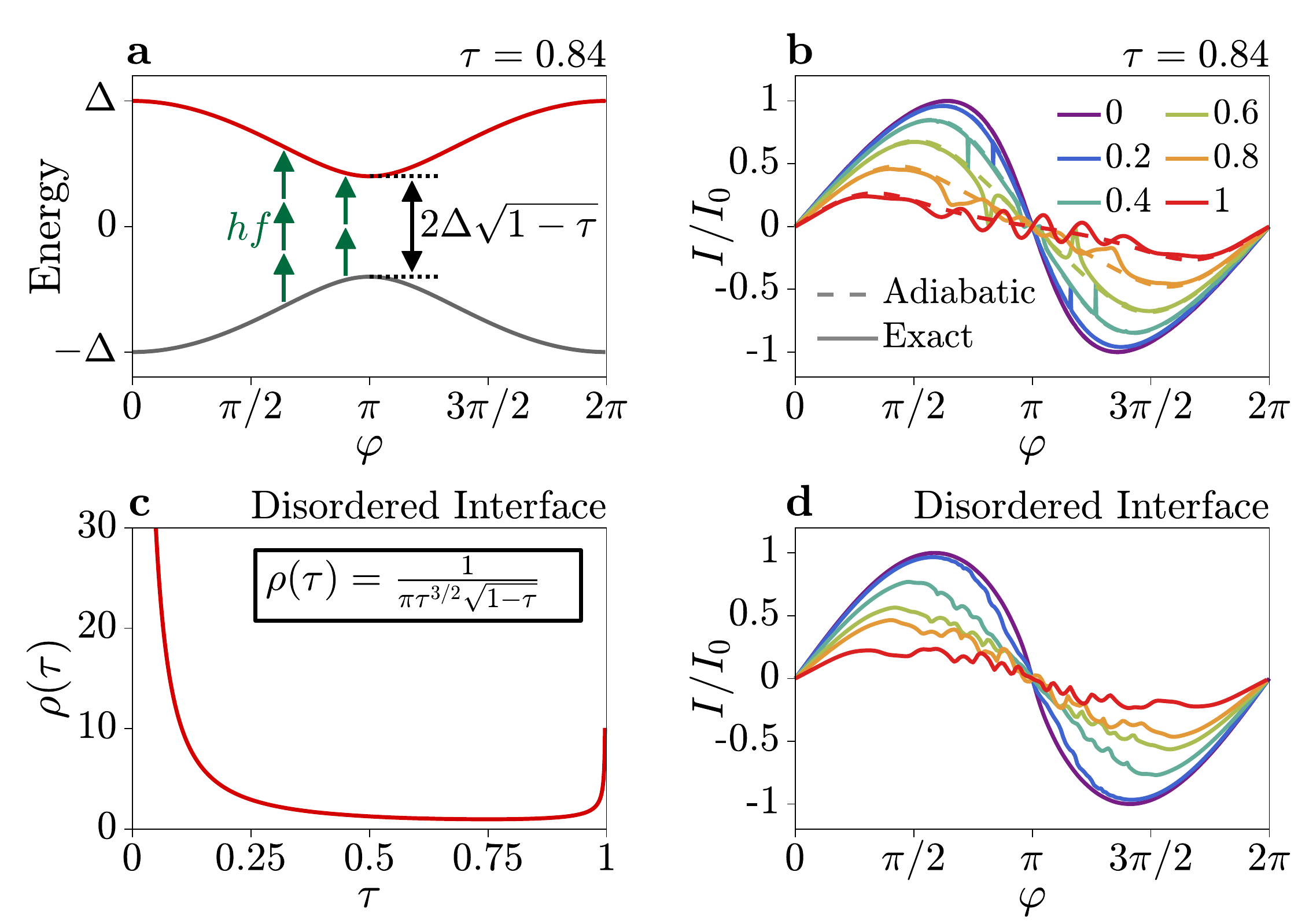}
	\caption{\textbf{Non-thermal occupation of Andreev bound states (ABSs).} \textbf{(a)} ABS spectrum for transmission $\tau=0.84$. Transitions (green) from occupied (grey) to unoccupied (red) states can occur close to $\varphi=\mathrm{\pi}$ by absorption of microwave photons with energy $hf$. \textbf{(b)} Normalised current-phase relation for transmission $\tau=0.84$ under increasing microwave field amplitude $\alpha$. Deviations of exact model (solid line) from adiabatic theory (dashed lines) occur at some values of $\varphi$ due to non-thermal occupation. \textbf{(c)} Distribution of channel transmissions for a planar Josephson junction with a disordered interface. The transmission distribution follows the equation in inset~\cite{Schep1997}. \textbf{(d)} Normalised current-phase relation for a junction modelled with a disordered interface, under increasing microwave field amplitude $\alpha$ [colour defined in (b)].}
	\label{Sfig10}
\end{figure}

At large applied microwave power, the measured CPR deviated from the fitted curve using the adiabatic model. At some values of the perpendicular magnetic field $\Bperp$, corresponding to certain phase values $\varphi$, the measured switching current was closer to zero than expected from the adiabatic model. This is interpreted as a non-thermal occupation of ABSs in the SNS junction, due to excitations driven by the microwave field. A microwave photon can induce a transition when the excitation energy $2\Ea$ is an integer multiple of the photon energy $hf$. Since $E_{\mathrm{A}}$ depends on the phase difference $\varphi$, absorption is expected only at specific $\varphi$ for a given frequency $f$. This is schematically shown in Fig.~\ref{Sfig10}(a), for the case of $\tau=0.84$. The current carried by an excited ABS is equal and opposite to that in the ground state, resulting in a suppression in the average measured current. For large drive powers, multi-photon processes are possible, and transitions can occur into or out of ABSs from the quasiparticle continuum. Excitation is most likely to occur close to $\varphi=\mathrm{\pi}$, since this is where $2\Ea$ is minimised. This is particularly true for highly transmissive ABSs, where the separation of the ABS from the superconducting gap edge can be large. To describe the impact of these different microwave-induced transitions on the CPR, we employed the theory of Ref.~\cite{Bergeret2011,Virtanen2010}. This theory, which is based on non-equilibrium Green's functions techniques, describes the CPR of a single channel superconducting point contact for arbitrary junction transparency ($\tau$) and strength of the coupling between the microwave field and the Josephson current ($\alpha=\mathrm{e}\Vmw/hf$). Figure~\ref{Sfig10}(b) shows the simulated CPR for microwave irradiation of $hf=0.19\Delta$, corresponding to a frequency of $9.20~\mathrm{GHz}$, for increasing $\alpha$ up to 1. The full model (solid lines) deviates from the adiabatic theory (dashed lines) for $\alpha\gtrapprox0.6$, consistent with the experimental observation.

The simulated CPR considers transitions in a single mode of transmission $\tau=0.84$, equal to the effective transmission of the junction. However, this does not consider the many modes present in the junction. Figure~\ref{Sfig10}(c) shows a distribution of transmissions in an SNS junction with a disordered interface, following the relation $\rho(\tau)=1/\pi\tau^{3/2}\sqrt{1-\tau}$~\cite{Schep1997}. The transmission distribution was chosen to give a CPR at equilibrium which matched the experimental result. The evolution of the CPR under microwave irradiation is shown in Fig.~\ref{Sfig10}(d). The suppression in switching current is less pronounced than in the single mode case, but occurs across a wider range of $\varphi$. The experimental data shows strong suppression across a wide range of $\varphi$, suggesting that the SNS junction is between the two extremes outlined in Fig.~\ref{Sfig10}. This is consistent with a junction containing many modes, some of which have a high transmission.

\bibliography{Bibliography2}

\end{document}